\documentclass[final,sort&compress,3p,times]{article}

\usepackage[utf8]{inputenc} 
\usepackage[T1]{fontenc}    
\usepackage{hyperref}       
\usepackage{url}            
\usepackage{booktabs}       
\usepackage{amsfonts}       
\usepackage{nicefrac}       
\usepackage{microtype}      
\usepackage{lipsum}
\usepackage{graphicx}

\usepackage{amssymb}
\usepackage{amsmath}
\usepackage{lineno}
\usepackage{caption}
\usepackage{subcaption}
\usepackage{setspace}

\usepackage{graphics}
\usepackage{ulem}
\usepackage{float}
\usepackage{tabu}
\usepackage{multirow}
\usepackage{pifont}
\usepackage{color}
\usepackage{authblk}

\title{Analysis of heat transfer and water flow with phase change in saturated porous media by bond-based peridynamics}

\date{}

\author[a]{\sffamily \small Petr Nikolaev \thanks{Corresponding author: petr.nikolaev@postgrad.manchester.ac.uk}}
\author[a]{\sffamily \small Majid Sedighi \thanks{Corresponding author: majid.sedighi@manchester.ac.uk}}
\author[a]{\sffamily \small Andrey P Jivkov}
\author[a]{\sffamily \small Lee Margetts }

\affil[a]{\scriptsize   Department of Mechanical, Aerospace and Civil Engineering, School of Engineering, The University of Manchester, United Kingdom}

\begin{document}
\maketitle
\begin{abstract}
A wide range of natural and industrial processes involve heat and mass transport in porous media. In some important cases the transported substance may undergo phase change, e.g. from liquid to solid and vice versa in the case of freezing and thawing of soils. The predictive modelling of such phenomena faces physical (multiple physical processes taking place) and mathematical (evolving interface with step change of properties) challenges. In this work, we develop and test a non-local approach based on bond-based peridynamics which addresses the challenges successfully. Our formulation allows for predicting the location of the interface between phases, and for calculating the temperature and pressure distributions within the saturated porous medium under the conditions of pressure driven water flow. The formulation is verified against existing analytical solutions for 1D problems, as well as finite element transient solutions for 2D problems. The agreement found by the verification exercise demonstrates the accuracy of the proposed methodology. The detailed coupled description of heat and hydraulic processes can be considered as a critical step towards a thermo-hydro-mechanical model, which will allow, for example, description of the hydrological behaviour of permafrost soils and the frost heave phenomenon.
\par\textbf{Keywords: } heat conduction; coupled problem; phase change;   peridynamics;  freezing; melting; porous medium
\end{abstract}

\section{Introduction}\label{sec:intro}
Heat and mass transfer with phase change occur in many natural and engineered solids. A non-exhaustive list of examples include seasonal and artificial freezing of soils, casting of metals, polymers and ceramics, and latent heat thermal energy storage.

Our focus is on porous media, with specific application to soils. In such systems, the phenomenon emerges from the coupling the multi-phase heat transfer with fluid flow through a porous substance. This involves strong physical non-linearities (rapid change of physical and mechanical properties between phases) and geometric discontinuities (inter-phase boundaries, cracks), making the mathematical description a challenging task. The lack of general analytical solutions \cite{Hahn2012HeatConduction} calls for appropriate numerical methods with specific attention to the strong non-linearities. To date, this problem has been approached by numerical methods based on the local (differential) formulation of the processes involved, e.g. by the finite element method (FEM). Such formulations are used in many engineering applications, for example in the analysis of: underground water flow in permafrost regions \cite{Rowland2011ThePermafrost, Orgogozo2019}; seasonally frozen soils \cite{Wan2021, ZHANG2021125603}; evolution of methane hydrates under the seabed \cite{Frederick2014TaliksConditions}; nuclear waste disposal design \cite{Vidstrand2013ModelingSweden, Grenier2013ImpactSystem}; and artificial ground freezing for mining and civil engineering purposes \cite{Vitel2016, Zhou2013, Tounsi2020Thermo-hydro-mechanicalFluid, Zueter2020ThermalModeling}.

However, accounting for the strong non-linearities and discontinuities is challenging for methods that are based on local mathematical formulations. The continued growth of computational power opens the opportunity to address these challenges by using methods based on non-local formulations, which are considered more computationally demanding. One non-local approach with increasing popularity is the Peridynamics (PD). In PD, the partial differential equations of any classical local theory are replaced by a set of integral-differential equations \cite{Madenci2014PeridynamicApplications}, resulting in a mathematically consistent formulation, even in the presence of strong non-linearities and discontinuities.

Peridynamics was originally proposed for describing mechanical behavior of solids \cite{Silling2000ReformulationForces, silling2007peridynamic}, and subsequently extended to a variety of diffusion problems \cite{Bobaru2012ADiscontinuities, Madenci2014PeridynamicApplications, Javili2019PeridynamicsReview}. It was first developed as a bond-based PD, and later generalized to a state-based PD in two versions - ordinary and non-ordinary \cite{Agwai2011, Madenci2014PeridynamicApplications}. Recently, an element-based PD formulation was proposed \cite{Liu2020APeridynamics}. A correspondence between PD and continuum formulations is made by the concept of the peridynamics differential operator (PDDO) \cite{Madenci2016, Madenci2019}. 

In this paper, we construct a bond-based PD model for transient heat transfer coupled with water flow and phase change in saturated porous media. The first PD formulation of heat conduction, which also considered electromigration, was made in \cite{gerstle2008peridynamic}. Bond-based formulations of transient heat conduction were originally developed in \cite{Bobaru2010TheConduction, Bobaru2012ADiscontinuities}, see also \cite{Chen2015SelectingDiffusion, Gu2019}. Further developments of the bond-based PD heat transfer models were discussed in \cite{Wang2016TheGreen, Jafari2017, Wang2018}. However, to date, the phase change phenomenon has not been included in the developments by this approach. In \cite{Madenci2017}, the one-dimensional phase change problem was solved using the PDDO framework, which prompted us to develop a bond-based PD model for obtaining such solutions when coupled with fluid flow. 

A PD formulation of pressure driven water flow in porous media was originally discussed in \cite{Katiyar2014AMedia, Jabakhanji2015AMedia}. This approach was recently used to develop a wide range of coupled problems that consider chemical diffusion in partially and fully saturated porous media e.g. \cite{Zhao2018ConstructionProblems, Sedighi2020PeridynamicsErosion, Yan2020PeridynamicsMedia, CHEN2021107463}. These demonstrated the simplicity and universality of the approach for dealing with such problems. However, the process of thermal diffusion has not yet been coupled with water flow in the analysis of heat and mass transfer in saturated porous media.

Our development solves the outstanding issues of coupling between different processes, following ideas from \cite{Zhao2018ConstructionProblems, Yan2020PeridynamicsMedia}. The present paper is structured as follows. In Section \ref{sec:local} we consider the classical local differential equations of heat transfer with phase change under pressure driven water flow in saturated porous media. In Section \ref{sec:peri}, we describe the basic concepts of bond-based peridynamics, and extend them to the case of heat transfer with phase change. Here, the formulation is also coupled with the existing model for pressure driven water flow. Section \ref{sec:numerical} presents our numerical implementation of the developed PD model for solution of 1D and 2D problems on regular square meshes. In Section \ref{sec:verify}, we consider several test problems and compare the solutions based on our model with existing analytical (1D) and numerical (2D by finite elements) solutions. The agreement between results demonstrates the accuracy of the proposed methodology. Importantly, our model is also successfully tested for 2D heat transfer with phase change with water flow due to high pressure gradients, a condition that is challenging for other methods including finite elements. Conclusions are drawn in Section \ref{sec:conclusions}.

The developed PD approach can be coupled with mechanical models that describe, for example, ocean ice cracking \cite{XUE2020107853, Lu2020Peridynamic, LIU2021108504}, cracking of concrete structures and rocks \cite{CHEN2021107463, ZHAO2020106969, CHEN2019104059}, as well as corrosion pitting \cite{CHEN2021104203}.

\section{Thermo-hydraulic model of porous medium with phase change}\label{sec:local}
In this study we consider a fully saturated porous medium with water as the pore-filling liquid. However, the considerations are applicable to any Newtonian fluid. The porous medium consists of three phases - solid matrix particles, liquid water, and solid water (ice). The physical parameters of these phases will be presented with lower indices $s$, $w$, and $i$, respectively. The hydraulic behavior of the liquid phase is governed by the mass conservation law, and the thermal behavior of the three-component structure is governed by the energy conservation law. The equations describing these laws can be constructed in the framework of the volume average method, see, for example, \cite{Vitel2016}. The equations in our presentation are written with respect to a Cartesian coordinate system.

The conservation of water mass in the two phases can be written as
\begin{equation} \label{eq:CEWM} 
    \frac{\partial}{\partial t} \left(n \rho_w S_w + n \rho_i S_i \right) = - \nabla\cdot \left(\rho_w \mathbf{u}\right),
\end{equation} 
where $S(\mathbf{x},t)$ is the saturation degree, for which $S_w +S_i =1$; $n$ is the porosity; $\rho$ is the density; $\mathbf{u}(\mathbf{x},t)$ is the flux of liquid water; $\mathbf{x}$ is the coordinate vector of a material point; and $t$ is the time.

In this work we assume a non-deformable pore system, so that the porosity $n$ is constant. The compressibility of ice is much lower than the compressibility of water, allowing us to consider the ice density to be approximately constant. With these assumptions, and taking into account that $S_w +S_i =1$, Eq. (\ref{eq:CEWM}) takes the form:
\begin{equation} \label{eq:CEWM_const_par} 
    n S_w \frac{\partial \rho_w}{\partial t} + n \left(\rho_w -\rho_i \right)\frac{\partial S_w}{\partial t}  = - \nabla \cdot \left(\rho_w \mathbf{u}\right). 
\end{equation} 

The water flow is described by Darcy's law:
\begin{equation} \label{eq:DL} 
    \mathbf{u}(\mathbf{x},t)=-\frac{k_r k_{int}}{\mu } \nabla \Phi,
\end{equation} 
where $k_r(\mathbf{x},t)$ is the liquid water relative permeability; $k_{int}$ is the porous medium intrinsic permeability; $\mu$ is the liquid water dynamic viscosity; and $\Phi (\mathbf{x},t)$ is the flow potential, which in terms of fluid pressure is given by:
\begin{equation} \label{eq:flow_pot} 
    \Phi (\mathbf{x},t)=p + g \rho_w z,
\end{equation} 
where $p(\mathbf{x},t)$ is the pressure; $g$ is the gravitational acceleration and $z(\mathbf{x},t)$ is the height of a liquid column. A number of relations have been proposed between the relative permeability, $k_r$ and the liquid water content. Most of the common relations for the frozen soils are presented in \cite{Kurylyk_Watanabe_2013, TENG2021125885}. In this paper, the relative permeability $k_r$ is defined according to \cite{McKenzie2007GroundwaterBogs} by:
\begin{equation} \label{eq:rel_perm} 
    k_r =10^{- n \Omega \left(1 - S_w \right)},
\end{equation} 
where $\Omega$ is an empirical parameter.

Substitution of Eq. (\ref{eq:DL}) into Eq. (\ref{eq:CEWM_const_par}) yields:
\begin{equation} \label{eq:CEWM_final} 
n S_w \dfrac{\partial \rho_w}{\partial t}  + n \left(\rho_w -\rho_i \right)\frac{\partial S_w}{\partial t} = \frac{k_{int}}{\mu} \nabla \cdot \left(\rho_w k_r \nabla \Phi \right),
\end{equation} 
which is the complete equation for the conservation of  mass for water flow given by Eq. (\ref{eq:flow_pot}) and considering phase change.

The energy conservation involves heat conduction and heat convection by a moving medium with phase change. Following \cite{Vitel2016}, this is presented by
\begin{equation} \label{eq:HC} 
    \left(\rho C \right)_{eq} \frac{\partial T}{\partial t} = - \nabla \cdot \mathbf{q} - \mathbf{u} \cdot \nabla \left(\rho_w C_w n S_w T\right),
\end{equation} 
where $\mathbf{q}(\mathbf{x},t)$ is the heat flux due to heat conduction; $T(\mathbf{x},t)$ is the temperature; $C$ is the specific heat capacity at a constant pressure; and $\left(\rho C \right)_{eq}(\mathbf{x},t)$ is the equivalent heat capacity of the three-phase medium defined by
\begin{equation} \label{eq:heat_cap_eq} 
    \left(\rho C \right)_{eq} =n S_w \rho_w C_w + n \left(1 - S_w \right) \rho_i C_i + \left(1 - n\right)\rho_s C_s + n \rho_i L \frac{\partial S_w }{\partial T}, 
\end{equation} 
where $L$ is the latent heat of water solidification. 

The dependence of the liquid water content on temperature below the freezing point is different for different materials. Specifically for soils, approximate relations for the soil freezing characteristic curves are discussed in \cite{YU2019150, WEN2020102927}. In the present study we consider a Weibull-type relation, used in \cite{McKenzie2007GroundwaterBogs}, according to which 
\begin{equation} \label{eq:water_saturation} 
S_w =\left\{\begin{array}{c} {1} \\ {(1-S_{w_{res}} )e^{-\left(\frac{T-T_f}{\Delta T} \right)^{2} } +S_{w_{res} } } \end{array}\quad \begin{array}{c} {T > T_f } \\ {T \le T_f} \end{array}\right.  
\end{equation} 
where $T_f$ is temperature at the onset of water solidification; $S_{w_{res}}$ is the residual saturation; $\Delta T$ is the Weibull scale parameter governing the smoothness of the $S_w$  function (the Weibul shape parameter is 2). This relation ensures fast and effective calculation. The derivative of $S_w$ with respect to temperature below freezing point is required for calculating the equivalent heat capacity in Eq. (\ref{eq:heat_cap_eq}). This is given by
\begin{equation} \label{eq:water_saturation_dir} 
    \frac{\partial S_w}{\partial T} = - 2 (1 - S_{w_{res}})\frac{T - T_f}{\Delta T^{2}} e^{-\left(\frac{T - T_f}{\Delta T} \right)^{2}}.
\end{equation} 

The heat flux due to the heat conduction is governed by Fourier's law:
\begin{equation} \label{eq:FL} 
    \mathbf{q}(x,t) = - \lambda\nabla T,
\end{equation} 
where $\lambda (\textbf{x}, t)$ is the material thermal conductivity. The thermal conductivity of the three-phase medium is given by
\begin{equation} \label{eq:heat_cond_mat} 
    \lambda = \lambda_w^{n S_w} \lambda_i^{n \left(1 - S_w \right)} \lambda_s^{\left(1 - n\right)}.
\end{equation} 

Substitution of Eqns. (\ref{eq:DL}), (\ref{eq:heat_cap_eq}) and (\ref{eq:FL}) into Eq. (\ref{eq:HC}) gives:
\begin{multline} \label{eq:HC_final} 
\left[n S_w \rho_w C_w + n \left(1 - S_w \right)\rho_i C_i +\left(1 - n \right)\rho_s C_s + n \rho_i L \frac{\partial S_w }{\partial T} \right]\frac{\partial T}{\partial t} = 
\\
-\nabla \left(\lambda \nabla T\right)+\rho_w C_w \nabla T \cdot \frac{k_r k_{int} }{\mu} \nabla \Phi,
\end{multline} 
which is the complete equation for the conservation of energy under flow potential given by Eq. (\ref{eq:flow_pot}) and considering phase change.

Equations (\ref{eq:CEWM_final}) and (\ref{eq:HC_final}) describe the processes of heat transfer and water flow with phase change in a local (differential) form. These will be reformulated in a non-local form in the next section.

\section{Bond-based peridynamics formulations for heat transfer with phase change and water flow} \label{sec:peri}
In bond-based Peridynamics, a body occupying a region $\mathbf{R}$ is considered as a collection of an arbitrary number of individual particles with associated volumes and masses. Each peridynamic particle is labeled by a position vector $\mathbf{x}$ with respect to a fixed (background) Cartesian coordinate system. A particle at position \textbf{x} interacts with (and is connected to) all particles at positions $\mathbf{x}'$ within a certain spatial region $\mathbf{H_x}$ referred to as the horizon of the particle at position $\textbf{x}$. The horizon may have any shape and size \cite{Gu2019}, and these may vary with the position $\mathbf{x}$, but in the majority of applications, as in this work, it is assumed to be of the form $\mathbf{H_x} = \{ \textbf{x}' : ||\textbf{x}'-\textbf{x}|| \leq \delta \}$, where the constant $\delta$ is referred to as the horizon radius. 

With this setting, the term 'bond' refers to the interactions between two particles at spatial positions $\mathbf{x}$ and $\mathbf{x}' \in \mathbf{H_x}$. Considering the nature of the processes to be modelled, we will refer to the bonds as transport or 't-bonds'. The peridynamic heat and fluid flux densities (fluxes per unit volume) along a 't-bond' depend on the distance between points $\textbf{x}$ and $\textbf{x'}$, represented by the distance vector $\mathbf{\xi} =\left(\mathbf{x'} - \mathbf{x}\right)$. 

\begin{figure}[ht]
    \centering
    \includegraphics[width=0.8\textwidth]{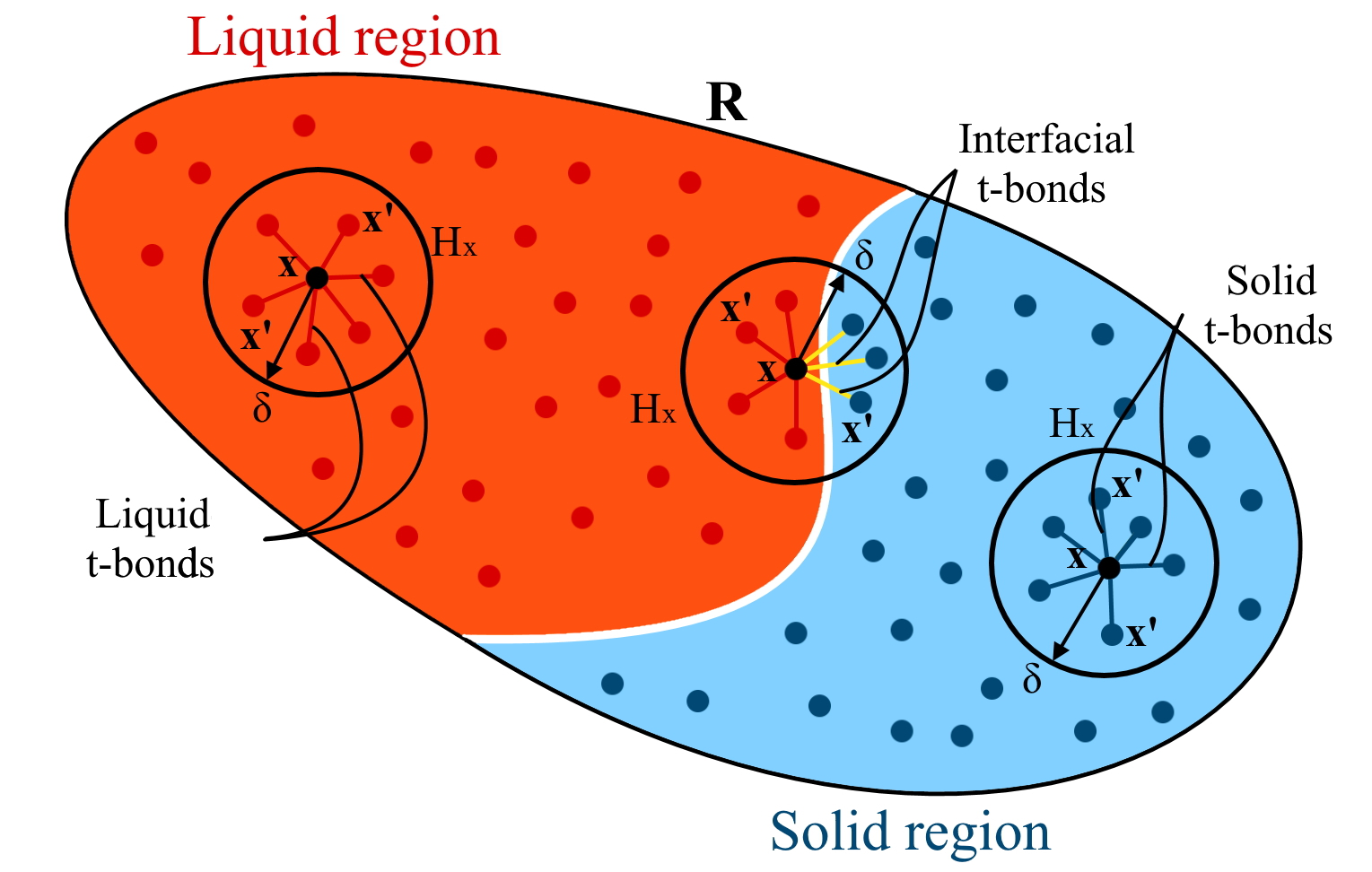}
    \caption{Illustration of phase regions, particle horizons and transport bonds (‘t-bonds’)}
    \label{fig:PD_model}
\end{figure}

With respect to the water phases, the region $\mathbf{R}$ can be divided into two regions -- solid and liquid -- as illustrated in Fig. \ref{fig:PD_model}. The solid region contains solid water (ice), and it may also contain the liquid water, which amount is defined by the soil freezing characteristic curve (\ref{eq:water_saturation}). The liquid region does not contain the solid water. The regions are separated by the surface with the temperature  $T_f$.

Depending on the region, where connected particles are located, we define three different PD t-bonds: solid t-bonds, interfacial t-bonds connecting different regions, and liquid t-bonds, see Fig. \ref{fig:PD_model}. If it is necessary, their transport properties can be considered differently. Thus, for a porous material, in which the solid region does not contain  liquid water or its transfer can be neglected, due to the very low relative permeability associated with it (in the present study defined by relation (\ref{eq:rel_perm})), it is suitable to assume that the solid t-bonds transport only heat, and do not participate in liquid transfer. Such bonds can be called 'impermeable'. The liquid and interfacial t-bonds transfer both heat and water. The conditions that attribute t-bonds to the 'impermeable' type can be chosen based on the analysis of considered problem. The t-bonds classification may be used to improve efficiency of the developed computational codes by not considering water flow through impermeable t-bonds.

\subsection{PD formulation of water flow}\label{sec:peri_flow}
The PD fluid flux density (flux per unit volume) along a t-bond between particles at positions $\mathbf{x}$ and $\mathbf{x}'$ per unit volume of $\mathbf{x}'$ is described by \cite{Katiyar2014AMedia}:
\begin{equation} \label{eq:PD_fludi_flux} 
    \overline{u}\left(\mathbf{x}',\mathbf{x},t\right)=\frac{k_{int} \overline{k_r }\left(\mathbf{x}',\mathbf{x},t\right)}{\mu } \frac{\Phi \left(\mathbf{x}',t\right)-\Phi \left(\mathbf{x},t\right)}{\left\| \xi \right\|} \frac{\xi }{\left\| \xi \right\| }, 
\end{equation}
where  $\overline{u}\left(\mathbf{x}',\mathbf{x},t\right)$ is the volumetric flux of liquid water; $\Phi \left(\mathbf{x},t\right)$ and $\Phi \left(\mathbf{x}',t\right)$ are flow potentials at points $\mathbf{x}$ and $\mathbf{x}'$, respectively;  $\mathbf{\xi} / \left\| \mathbf{\xi} \right\|$ is the unit vector along the t-bond; and $\overline{k_r }\left(\mathbf{x}',\mathbf{x},t\right)$ is the relative permeability of the t-bond between the points $\mathbf{x}$ and $\mathbf{x}'$.

Since the t-bond relative permeability depends on the locations of its end points within the two different regions, which properties can vary significantly, we define it in the following way:
\begin{equation} \label{eq:DB_relat_perm} 
    \overline{k_r }\left(\mathbf{x}',\mathbf{x},t\right)=\frac{2k_r\left(\mathbf{x}',t\right)k_r \left(\mathbf{x},t\right)}{k_r\left(\mathbf{x}',t\right)+k_r\left(\mathbf{x},t\right)} 
\end{equation} 
This definition ensures that, the solid and interfacial t-bonds would have  $\overline{k_r }\left(\mathbf{x}',\mathbf{x},t\right)\approx 0$, which cannot be achieved with a simple average. 

The mass conservation equation for a t-bond between points $\mathbf{x}$ and $\mathbf{x}'$ is given by \cite{Katiyar2014AMedia, Yan2020PeridynamicsMedia}
\begin{equation} \label{eq:PD_mass_cons} 
    n S_w \frac{\partial \rho_w \left(\mathbf{x}',\mathbf{x},t\right)}{\partial t} +R_b \left(\mathbf{x}',\mathbf{x}\right)=\frac{k_{int} \overline{k_r }\left(\mathbf{x}',\mathbf{x},t\right) }{\mu } \rho_w \frac{\Phi \left(\mathbf{x'},t\right)-\Phi \left(\mathbf{x},t\right)}{\left\| \xi \right\| ^{2}},  
\end{equation} 
where $R_{b}\left(\mathbf{x}',\mathbf{x}'\right)$ is a volumetric mass source term, that can be represented by
\begin{equation} \label{ew:vol_mass_source} 
R_{b}\left(\mathbf{x}',\mathbf{x},t\right)=n\left(\rho_w -\rho_i \right)\frac{\partial S_w \left(\mathbf{x}',\mathbf{x},t\right)}{\partial t}.
\end{equation} 

The mass conservation at point $\mathbf{x}$ is obtained by integrating Eq. (\ref{eq:PD_mass_cons}) over its horizon $\mathbf{H_x}$
\begin{multline} \label{eq:PD_mass_cons_1} 
\int_{\mathbf{H_x}}nS_w \frac{\partial \rho_w  \left(\mathbf{x}',\mathbf{x},t\right)}{\partial t} {\rm d}V_{\mathbf{x}'} +\int _{\mathbf{H_x}}n\left(\rho_w -\rho_i \right)\frac{\partial S_w \left(\mathbf{x}',\mathbf{x},t\right)}{\partial t}  {\rm d}V_{\mathbf{x}'} = 
\\
\int_{\mathbf{H_x}} \frac{k_{int} \overline{k_r }\left(\mathbf{x}',\mathbf{x},t\right) }{\mu } \rho_w \frac{\Phi \left(\mathbf{x}',t\right)-\Phi \left(\mathbf{x},t\right)}{\left\| \xi \right\| ^{2} } {\rm d}V_{\mathbf{x}'} 
\end{multline} 

The relation between the water density at point $\mathbf{x}$ and time $t$ and the water density in all adjacent t-bonds is given by \cite{Yan2020PeridynamicsMedia}
\begin{equation} \label{eq:PD_mass_cons_2} 
\int_{\mathbf{H_x}}n S_w \frac{\partial \rho_w \left(\mathbf{x}',\mathbf{x},t\right)}{\partial t}  {\rm d}V_{\mathbf{x}'} =nS_w \frac{\partial \rho_w \left(\mathbf{x},t\right)}{\partial t} V_{\mathbf{x} }
\end{equation} 

Similarly, according to \cite{Katiyar2014AMedia}, the mass source term at point $\mathbf{x}$ can be obtained as the average in all adjacent t-bonds:
\begin{equation} \label{eq:PD_mass_cons_3} 
\int_{\mathbf{H_x}}n\left(\rho_w -\rho_i \right)\frac{\partial S_w \left(\mathbf{x}',\mathbf{x},t\right)}{\partial t}  {\rm d}V_{\mathbf{x}'} =n\left(\rho_w -\rho_i \right)\frac{\partial S_w \left(\mathbf{x},t\right)}{\partial t} V_{\mathbf{x}}  
\end{equation} 

Combining Eqns. (\ref{eq:PD_mass_cons}) -- (\ref{eq:PD_mass_cons_3}) leads to the mass conservation equation for liquid water at point $\mathbf{x}$ in terms of density and flow potential 
\begin{multline} \label{eq:PD_mass_cons_4} 
nS_w \frac{\partial }{\partial t} \rho_w \left(\mathbf{x},t\right)+n\left(\rho_w -\rho_i \right)\frac{\partial }{\partial t} S_w \left(\mathbf{x},t\right)= \\
\int_{\mathbf{H_x}}\frac{\rho_w }{\mu } \frac{k_{int} \overline{k_r }\left(\mathbf{x}',\mathbf{x},t\right)}{V_{\mathbf{x}} } \frac{\Phi \left(\mathbf{x}',t\right)-\Phi \left(\mathbf{x},t\right)}{\left\| \xi \right\| ^{2} }  {\rm d}V_{\mathbf{x}'}  
\end{multline} 
Under the assumption that $\rho_w \left(\mathbf{x},t\right)=const$, the last equation becomes
\begin{equation} \label{eq:PD_mass_cons_5} 
n\left(\rho_w -\rho_i \right)\frac{\partial }{\partial t} S_w \left(\mathbf{x},t\right)=
\int_{\mathbf{H_x}}\frac{k_{int} \overline{k_r }\left(\mathbf{x}',\mathbf{x},t\right)}{V_{\mathbf{x}} } \frac{\rho_w }{\mu } \frac{\Phi \left(\mathbf{x}',t\right)-\Phi \left(\mathbf{x},t\right)}{\left\| \xi \right\| ^{2} }  {\rm d}V_{\mathbf{x}'}  
\end{equation} 
The PD micro-permeability of a t-bond can be defined as:
\begin{equation} \label{eq:PD_micro_perm} 
K_{w}\left(\mathbf{x}',\mathbf{x},t\right)=\frac{k_{int} \overline{k_r }\left(\mathbf{x}',\mathbf{x},t\right)}{\mu V_{\mathbf{x} } }  
\end{equation} 

Using the approach of uniform constant influence functions \cite{Zhao2018ConstructionProblems} the relations between PD micro-hydraulic conductivity $K_{w}\left(\mathbf{x}',\mathbf{x},t\right)$ and the macroscopic properties for 1D and 2D cases is derived as \cite{Yan2020PeridynamicsMedia}
\begin{equation} \label{eq:PD_micro_perm_1)} 
K_{w}\left(\mathbf{x}',\mathbf{x},t\right) = \frac{k_{int} \overline{k_r } \left(\mathbf{x}',\mathbf{x},t\right)}{\mu \delta }  
\end{equation} 
\begin{equation} \label{eq:PD_micro_perm_2)} 
K_{w}\left(\mathbf{x}',\mathbf{x},t\right)=\frac{4k_{int} \overline{k_r } \left(\mathbf{x}',\mathbf{x},t\right)}{\mu \pi \delta ^{2}}  
\end{equation}

For the known pressure field, that can be found by solution of the equation (\ref{eq:PD_mass_cons_4}), following  \cite{Katiyar2014AMedia}, we can define the water flux at any point $\textbf{x}$ by the relation:
\begin{equation}
\label{eq:water_flux} 
\mathbf{u}(\mathbf{x}, t) = \int_{\mathbf{H_x}} K_{w}\left(\mathbf{x}',\mathbf{x},t\right) \rho_w  \frac{\Phi \left(\mathbf{x}',t\right)-\Phi \left(\mathbf{x},t\right)}{\left\| \xi \right\|} \frac{\mathbf{\xi}}{\left\| \mathbf{\xi} \right\|} {\rm d}V_{\mathbf{x}'}  
\end{equation} 
For the 2-D case, the vector $\mathbf{u}(\mathbf{x}, t)$  can be considered as two components: $u_x(\mathbf{x}, t)$ and $u_y(\mathbf{x}, t)$ along the $x$ and $y$ axis, respectively.

\subsection{PD formulation of heat transfer with phase change}\label{sec:peri_heat}
The heat flux along a t-bond, $J\left(T,\mathbf{x}',\mathbf{x},t\right)$, can be defined according to \cite{Zhao2018ConstructionProblems, Yan2020PeridynamicsMedia} as:
\begin{multline} \label{eq:DB_heat_flux} 
    J = \overline{\lambda}\left(\mathbf{x}', \mathbf{x}, t\right) \frac{T\left(\mathbf{x}',t\right) - T\left(\mathbf{x}, t\right)}{\left\| \mathbf{\xi} \right\|} \cdot \frac{\mathbf{\xi}}{\left\| \mathbf{\xi} \right\|} - \\ \rho_w C_w  \overline{u}\left(\mathbf{x}', \mathbf{x}, t\right) \left[T\left(\mathbf{x}', t\right) -  T\left(\mathbf{x}, t\right)\right] \cdot \frac{\mathbf{\xi}}{\left\| \mathbf{\xi} \right\|}
\end{multline} 
where $T\left(\mathbf{x}', t\right)$ and $T\left(\mathbf{x}, t\right)$ are the temperature at points $\mathbf{x}'$ and $\mathbf{x}$ at time $t$, respectively; $\overline{\lambda}\left(\mathbf{x}', \mathbf{x},t\right)$ is the thermal conductivity of the t-bond; and $\overline{u}\left(\mathbf{x}',\mathbf{x},t\right)$ is the average water flux along the t-bond.

The thermal conductivity of a t-bond depends on the locations of the two connected particles in the three considered regions, and we define it with a simple interpolation by
\begin{equation} \label{eq:DB_heat_cond)} 
    \overline{\lambda }\left(\mathbf{x}', \mathbf{x}, t\right)=\frac{\lambda (\mathbf{x},t)+\lambda (\mathbf{x}',t)}{2},
\end{equation} 
where $\lambda (\mathbf{x},t)$ and $\lambda (\mathbf{x}',t)$ are the thermal conductivities at the peridynamic particles located at $\mathbf{x}$ and $\mathbf{x}'$, respectively.

Similarly, and following \cite{Yan2020PeridynamicsMedia, Oterkus2017FullyFractures}, the water velocity along a t-bond is defined by 
\begin{equation} \label{eq:DB_water_flux} 
    \overline{u}\left(\mathbf{x}', \mathbf{x}, t\right)=\frac{\mathbf{u}\left(\mathbf{x}, t\right) + \mathbf{u}\left(\mathbf{x}',t\right)}{2},  
\end{equation} 
where the liquid velocity vectors $\mathbf{u}\left(\mathbf{x}, t\right)$ and $\mathbf{u}\left(\mathbf{x}',t\right)$ at particles at positions $\mathbf{x}$ and $\mathbf{x}'$, respectively, are found by solving the equation (\ref{eq:water_flux}) for a given pressure field. This relation ensures that both liquid and inetrfacial types of t-bonds participate in convective heat transfer.

The peridynamic version of the energy conservation equation is written for each t-bond in the following form \cite{Bobaru2012ADiscontinuities, Bobaru2010TheConduction, Chen2015SelectingDiffusion}
\begin{equation} \label{eq:DB_energy_cons} 
    \overline{\left(\rho C\right)_{eq} }(\mathbf{x}', \mathbf{x}, t)\frac{\partial }{\partial t} \overline{T}(\mathbf{x}', \mathbf{x}, t)=\frac{J\left(\mathbf{x}', \mathbf{x}, t\right)}{\left\| \xi \right\|} 
\end{equation} 
where $\overline{T}(\mathbf{x}', \mathbf{x}, t)$ is the t-bond temperature taken as the average of the temperatures at points $\mathbf{x}$ and $\mathbf{x}'$, and $\overline{\left(\rho C\right)_{eq} }(\mathbf{x}', \mathbf{x}, t)$ is the t-bond equivalent heat capacity taken as the average of the equivalent heat capacities at points $\mathbf{x}$ and $\mathbf{x}'$.

Inserting Eq. (\ref{eq:DB_heat_flux}) into Eq. (\ref{eq:DB_energy_cons}) provides the energy conservation in a t-bond in terms of temperature and water velocity
\begin{multline} \label{eq:DB_energy_cons_2)} 
    \overline{\left(\rho C\right)_{eq} }(\mathbf{x}',\mathbf{x},t)\frac{\partial \overline{T}(\mathbf{x}',\mathbf{x},t)}{\partial t} =
    \overline{\lambda }\left(\mathbf{x}',\mathbf{x},t\right)\frac{T\left(\mathbf{x}',t\right)-T\left(\mathbf{x},t\right)}{\left\| \xi \right\| ^{2} } \cdot \frac{\xi}{\left\| \xi \right\| } -
    \\
    \rho_w C_w \overline{u}\left(\mathbf{x}',\mathbf{x},t\right)\frac{T\left(\mathbf{x}',t\right)-T\left(\mathbf{x},t\right)}{\left\| \xi \right\| } \cdot \frac{\xi }{\left\| \xi \right\| } 
\end{multline} 

The energy conservation for a particle at position $\mathbf{x}$ involves the fluxes in all adjacent t-bonds (bonds to particles within the horizon $\mathbf{H_x}$) and is obtained by integrating Eq. (\ref{eq:DB_energy_cons}) over the horizon:
\begin{multline} \label{eq:DB_energy_cons_3)} 
    \int _{\mathbf{H_x} }\overline{\left(\rho C\right)_{eq} }(\mathbf{x}',\mathbf{x},t)\frac{\partial \overline{T}(\mathbf{x}',\mathbf{x},t)}{\partial t}  {\rm d}V_{\mathbf{x}'} = \\
    \int _{\mathbf{H_x} }  \overline{\lambda }\left(\mathbf{x}',\mathbf{x},t\right)\frac{T\left(\mathbf{x}',t\right)-T\left(\mathbf{x},t\right)}{\left\| \xi \right\| ^{2} } \cdot \frac{\xi }{\left\| \xi \right\| }  - 
    \\ 
    \rho_w C_w \overline{u}\left(\mathbf{x}',\mathbf{x},t\right)\frac{T\left(\mathbf{x}',t\right)-T\left(\mathbf{x},t\right)}{\left\| \xi \right\| } \cdot \frac{\xi }{\left\| \xi \right\|}] {\rm d}V_{\mathbf{x}'} 
\end{multline} 
where $V_{\mathbf{x}'}$ is the horizon volume of the particle at position $\mathbf{x}'$.

The relationship between temperature at point $\mathbf{x}$ and time $t$ and the temperature in all adjacent bonds can be written as \cite{Bobaru2010TheConduction}:
\begin{equation} \label{eq:DB_energy_cons_4} 
    \int _\mathbf{H_x}\overline{\left(\rho C\right)_{eq}}(\mathbf{x}',\mathbf{x},t)\frac{\partial \overline{T}(\mathbf{x}',\mathbf{x},t)}{\partial t}  {\rm d}V_{\mathbf{x}'} =\overline{\left(\rho C\right)_{eq} }(\mathbf{x},t)\frac{\partial T(\mathbf{x},t)}{\partial t} V_{\mathbf{x} }  
\end{equation} 
where $V_{\mathbf{x}}$ is the horizon volume of particle $\mathbf{x}$. Note, that under the assumption of constant horizon radius $V_{\mathbf{x}'} = V_{\mathbf{x}}$, and specifically for 1D problems $V_{\mathbf{x}} = 2\delta$, and for 2D problems $V_{\mathbf{x}} = \pi \delta^2$.

Combining Eqns. (\ref{eq:DB_heat_flux}) -- (\ref{eq:DB_energy_cons_4}) leads to the following equation for the evolution of temperature at particle $\mathbf{x}$
\begin{multline} \label{eq:PD_temp_ev} 
    \overline{\left(\rho C\right)_{eq} }(\mathbf{x},t)\frac{\partial T(\mathbf{x},t)}{\partial t} =\int_{\mathbf{H_x}}\frac{\overline{\lambda }\left(\mathbf{x}',\mathbf{x},t\right)}{V_{\mathbf{x} } } \frac{T\left(\mathbf{x}',t\right)-T\left(\mathbf{x},t\right)}{\left\| \xi \right\| ^{2} } \cdot \frac{\xi }{\left\| \xi \right\| } -
    \\ 
     \rho_w C_w \frac{\overline{u}\left(\mathbf{x}',\mathbf{x},t\right)}{V_{\mathbf{x} } } \frac{T\left(\mathbf{x}',t\right)-T\left(\mathbf{x},t\right)}{\left\| \xi \right\| } \cdot \frac{\xi }{\left\| \xi \right\| } {\rm d}V_{\mathbf{x}'} 
\end{multline} 

The peridynamics microscopic heat conductivity and the peridynamics microscopic liquid velocity at particle $\mathbf{x}$ are defined by \cite{Zhao2018ConstructionProblems}:
\begin{equation} \label{eq:PD_diffusiv} 
    \Lambda\left(\mathbf{x}',\mathbf{x},t\right)=\frac{\overline{\lambda }\left(\mathbf{x}',\mathbf{x},t\right)}{V_{\mathbf{x} } } 
\end{equation} 
\begin{equation}\label{eq:PD_m_v_d} 
    U\left(\mathbf{x}',\mathbf{x},t\right)=\frac{\overline{u}\left(\mathbf{x'},\mathbf{x},t\right)}{V_{\mathbf{x}}},
\end{equation} 
Therefore, Eq. (\ref{eq:PD_temp_ev}) is written as:
\begin{multline} \label{eq:PD_temp_ev_1} 
    \overline{\left(\rho C\right)_{eq} }(\mathbf{x},t)\frac{\partial T(\mathbf{x},t)}{\partial t} =\int_{\mathbf{H_x}}\Lambda \left(\mathbf{x}',\mathbf{x},t\right)\frac{T\left(\mathbf{x}',t\right)-T\left(\mathbf{x},t\right)}{\left\| \xi \right\| ^{2} } -
    \\
     \rho_w C_w \left(\mathbf{x}',\mathbf{x},t\right) U \left(\mathbf{x}',\mathbf{x},t\right)\frac{T\left(\mathbf{x}',t\right)-T\left(\mathbf{x},t\right)}{\left\| \xi \right\| } {\rm d}V_{\mathbf{x}'}
\end{multline} 

Using the approach of uniform constant influence functions proposed in \cite{Bobaru2010TheConduction, Chen2015SelectingDiffusion}, the relationship between PD microscopic heat conductivity $\Lambda \left(\mathbf{x}',\mathbf{x},t\right)$ and macroscopic (which is experimentally measured) heat conductivity $\overline{\lambda}$ for the 1D and 2D cases can be derived by equating PD and classical local solutions. The results are
\begin{equation} \label{eq:PD_mic_liq_flux_1)} 
    \Lambda\left(\mathbf{x}',\mathbf{x},t\right)=\frac{\overline{\lambda } \left(\mathbf{x}',\mathbf{x},t\right)}{\delta }  
\end{equation} 
\begin{equation} \label{eq:PD_mic_liq_flux_2)} 
    \Lambda\left(\mathbf{x}',\mathbf{x},t\right)=\frac{4\overline{\lambda } \left(\mathbf{x}',\mathbf{x},t\right)}{\pi \delta ^{2} }  
\end{equation} 
Similarly, the relationship between PD microscopic water velocity $U \left(\mathbf{x}',\mathbf{x},t\right)$ and the macroscopic velocity $\overline{u}$ for the 1D and 2D cases are \cite{Zhao2018ConstructionProblems}:
\begin{equation} \label{eq:PD_mic_liq_flux_1} 
    U \left(\mathbf{x}',\mathbf{x},t\right)=\frac{\overline{u} \left(\mathbf{x}',\mathbf{x},t\right)}{2\delta}  
\end{equation} 
\begin{equation} \label{eq:PD_mic_liq_flux_2} 
    U \left(\mathbf{x}',\mathbf{x},t\right)=\frac{2\overline{u} \left(\mathbf{x}',\mathbf{x},t\right)}{\pi \delta^{2} }  
\end{equation}

\section{Numerical implementation}\label{sec:numerical}
Each peridynamic particle has an associated volume, which in the cases considered here is represented by length in 1D and area in 2D. In our implementation, we consider that all particles have equal volumes. In such case, the spatial discretisation of the domain is by a uniform grid with step equal to the particle size, $\Delta x$, which is the length in 1D and the square root of the area in 2D. The temporal discretisation is implemented by considering equally spaced time instances, $t^n$ ($n \in \mathbb{N}$), with time interval between the instances $\Delta t = t^{n+1} - t^n$. 

Since the pressure gradients in the majority of engineering problems discussed in Section \ref{sec:intro} are insignificant, we consider $\rho_w \left(\mathbf{x},t\right)=const$ and discretise Eq. (\ref{eq:PD_mass_cons_5}). The spatial discretisation of the conservation of mass in particle at position $\mathbf{x}_{\alpha}$ ($\alpha \in \mathbb{N}$) is written as
\begin{equation} \label{eq:PD_mass_space_disc} 
n\frac{\left(\rho_w -\rho_i \right)}{\rho_w } \frac{\partial S_w\left(\mathbf{x}_{\alpha} ,t\right)}{\partial t} =\sum_{\mathbf{x}_{\beta}} K_w\left(\mathbf{x}_{\beta} ,\mathbf{x}_{\alpha} ,t\right)\frac{\Phi\left(\mathbf{x}_{\beta} ,t\right)-\Phi\left(\mathbf{x}_{\alpha} ,t\right)}{\left\| \mathbf{x}_{\beta} -\mathbf{x}_{\alpha} \right\| ^{2} } V_{\alpha\beta}   
\end{equation} 
where $\mathbf{x}_{\beta}$ ($\beta \in \mathbb{N}$) are the positions of particles in the horizon of $\textbf{x}_{\alpha}$, $V_{\alpha\beta}$ is the portion of the volume associated with $\mathbf{x}_{\beta}$ within the horizon of $\mathbf{x}_{\alpha}$, and $K_w\left(\mathbf{x}_{\beta} ,\mathbf{x}_{\alpha} ,t\right)$ is calculated by Eq. (\ref{eq:DB_relat_perm}). We consider a horizontal water table, so that $\Phi (\mathbf{x},t)=p(\mathbf{x},t)$, see Eq. (\ref{eq:flow_pot}), for which Eq. (\ref{eq:PD_mass_space_disc}) becomes:
\begin{equation} \label{eq:PD_mass_space_disc_2} 
n\frac{\left(\rho_w -\rho_i \right)}{\rho_w } \frac{\partial S_w\left(\mathbf{x}_{\alpha} ,t\right)}{\partial t} =\sum_{\mathbf{x}_{\beta}} K_w\left(\mathbf{x}_{\beta} ,\mathbf{x}_{\alpha} ,t\right)\frac{p\left(\mathbf{x}_{\beta} ,t\right)-p\left(\mathbf{x}_{\alpha} ,t\right)}{\left\| \mathbf{x}_{\beta} -\mathbf{x}_{\alpha} \right\| ^{2} } V_{\alpha\beta}   
\end{equation} 
The time derivative of the water saturation in Eq. (\ref{eq:PD_mass_space_disc_2}) is calculated using the value from the previous time step, so that the fully discretised conservation of mass becomes
\begin{multline} \label{eq:PD_press_discret_time} 
\frac{S_w\left(\mathbf{x}_{\alpha},t^{n} \right)-S_w\left(\mathbf{x}_{\alpha} ,t^{n-1} \right)}{\Delta t} = \\
\sum _{p}\frac{K_{w} \left(\mathbf{x}_{\beta},\mathbf{x}_{\alpha},t^{n} \right)\rho_w }{n\left(\rho_w -\rho_i \right)} \frac{p\left(\mathbf{x}_{\beta} ,t^{n} \right)-p\left(\mathbf{x}_{\alpha},t^{n} \right)}{\left\| \mathbf{x}_{\beta} -\mathbf{x}_{\alpha} \right\| ^{2} } V_{\alpha\beta}   
\end{multline}

For the known pressure field, that is defined by the solution of a system of linear equations (\ref{eq:PD_press_discret_time}), we can define the vector of water flux $\mathbf{u}(\mathbf{x}, t)$ by the numerical solution of the relation (\ref{eq:water_flux}). It is suitable to find the components of this vector separately, for that, the following system can be solved:
\begin{equation} \label{eq:water_flux_disrect} 
\mathbf{u}(\mathbf{x}, t) = 
\begin{bmatrix}
u_x \\
u_y
\end{bmatrix}
=
\begin{bmatrix}
\sum _{p}\frac{K_{w} \left(\mathbf{x}_{\beta},\mathbf{x}_{\alpha},t^{n} \right)\rho_w }{n\left(\rho_w -\rho_i \right)} \frac{p\left(\mathbf{x}_{\beta} ,t^{n} \right)-p\left(\mathbf{x}_{\alpha},t^{n} \right)}{\left\| \mathbf{x}_{\beta} -\mathbf{x}_{\alpha} \right\| ^{2} } V_{\alpha\beta} \cdot cos(\gamma_{\alpha,\beta})  \\
\sum _{p}\frac{K_{w} \left(\mathbf{x}_{\beta},\mathbf{x}_{\alpha},t^{n} \right)\rho_w }{n\left(\rho_w -\rho_i \right)} \frac{p\left(\mathbf{x}_{\beta} ,t^{n} \right)-p\left(\mathbf{x}_{\alpha},t^{n} \right)}{\left\| \mathbf{x}_{\beta} -\mathbf{x}_{\alpha} \right\| ^{2} } V_{\alpha\beta} \cdot sin(\gamma_{\alpha,\beta})
\end{bmatrix}
\end{equation}
where $\gamma_{\alpha,\beta}$ is the angle between the unit vector along a considered t-bond and the unit vector $\textbf{j}$ that corresponds to horizontal axes $x$. 

After the water flux vectors $\mathbf{u}(\mathbf{x}, t)$ are defined for all particle within the considered domain $\mathbf{R}$, the average water velocity along every t-bond, that is necessary to consider the conservation of energy, can be calculated by the relation (\ref{eq:DB_water_flux}). It finally lets us define the microscopic liquid velocity according to (\ref{eq:PD_mic_liq_flux_1}) or (\ref{eq:PD_mic_liq_flux_2}).

The spatial discretisation of the conservation of energy, Eq. (\ref{eq:PD_temp_ev_1}), for a particle at position $\mathbf{x}_{\alpha}$ is
\begin{multline} \label{eq:PD_temp_discret} 
\overline{\left(\rho C\right)_{eq} }(\mathbf{x}_{\alpha} ,t)\frac{\partial T(\mathbf{x}_{\alpha} ,t)}{\partial t} =
\sum_{\mathbf{x}_{\beta}} \left[ \Lambda \left(\mathbf{x}_{\beta} ,\mathbf{x}_{\alpha} ,t\right)\frac{T\left(\mathbf{x}_{\beta} ,t\right)-T\left(\mathbf{x}_{\alpha} ,t\right)}{\left\| \mathbf{x}_{\beta} -\mathbf{x}_{\alpha} \right\| ^{2} } V_{\alpha\beta}  \right. - 
\\
\left. \rho_w C_w U \left(\mathbf{x}_{\beta},\mathbf{x}_{\alpha} ,t\right)\frac{T\left(\textbf{x}_{\beta} ,t\right)-T\left(\mathbf{x}_{\alpha},t\right)}{\left\| \mathbf{x}_{\beta} -\mathbf{x}_{\alpha} \right\| } V_{\alpha\beta} \right]
\end{multline} 
The time derivative of temperature in Eq. (\ref{eq:PD_temp_discret}) is approximated by the forward Euler method. Thus, the fully discretised conservation of energy becomes
\begin{multline} \label{eq:PD_temp_discret_time} 
\frac{T\left(\mathbf{x}_{\alpha},t^{n+1} \right)-T\left(\mathbf{x}_{\alpha},t^{n} \right)}{\Delta t} = \\
\sum _{p} \left[ \frac{\Lambda \left(\mathbf{x}_{\beta},\mathbf{x}_{\alpha},t^{n} \right)}{\left(\rho C\right)_{eq} \left(\mathbf{x}_{\alpha},t^{n} \right)} \frac{T\left(\mathbf{x}_{\beta},t^{n} \right)-T\left(\mathbf{x}_{\alpha},t^{n} \right)}{\left\| \mathbf{x}_{\beta} -\mathbf{x}_{\alpha} \right\| ^{2} } V_{\alpha\beta}  - \right. 
\\
  \left.\frac{\rho_w C_w U \left(\mathbf{x}_{\beta},\mathbf{x}_{\alpha},t^{n} \right)}{\left(\rho C\right)_{eq} \left(\mathbf{x}_{\alpha},t^{n} \right)} \frac{T\left(\mathbf{x}_{\beta},t^{n} \right)-T\left(\mathbf{x}_{\alpha},t^{n} \right)}{\left\| \mathbf{x}_{\beta} -\mathbf{x}_{\alpha} \right\| } V_{\alpha\beta} \right]
\end{multline}

\begin{figure}[ht]
    \centering
    \includegraphics[width=0.7\textwidth]{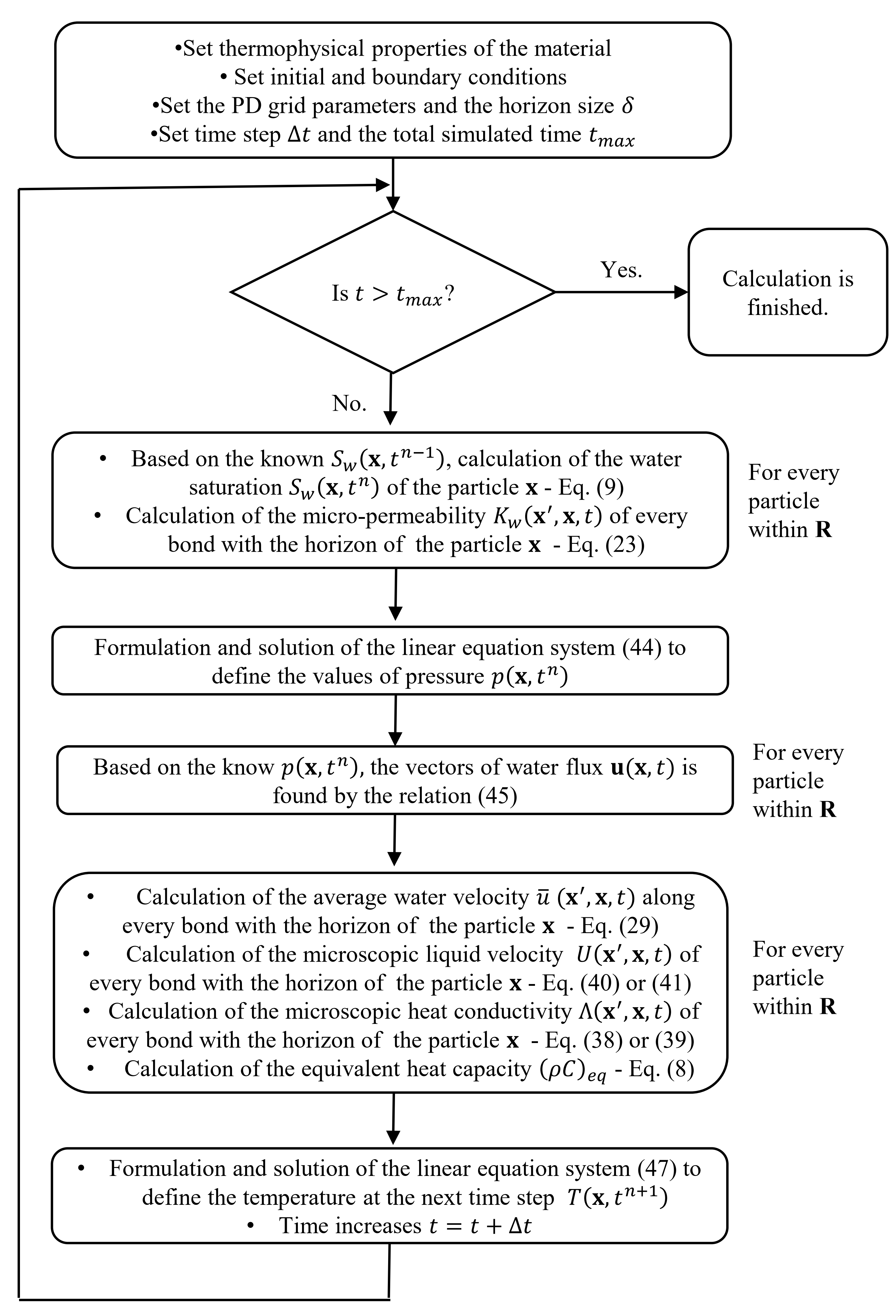}
    \caption{The general flow-chart of the developed PD model}
    \label{fig:PD_Flow_chart}
\end{figure}

Equations (\ref{eq:PD_press_discret_time}) and (\ref{eq:PD_temp_discret_time}) represent the numerical implementation of the developed bond-based PD model. Based on the known $S_w\left(\mathbf{x} ,t^{n-1}\right)$ and $S_w\left(\mathbf{x} ,t^{n}\right)$, the solution of Eq. (\ref{eq:PD_press_discret_time}) provides the pressure field at $t^n$, that defines the average water flux along every t-bond (\ref{eq:DB_water_flux}). Taking it into account, the vectors of water flux at every particle can be found by the relation (\ref{eq:water_flux_disrect}). Finally, the solution to Eq. (\ref{eq:PD_temp_discret_time}) provides the temperature field at $t^{n+1}$. The detailed algorithm of the proposed numerical implementation is presented in Fig . \ref{fig:PD_Flow_chart}.

According to \cite{Zhao2018ConstructionProblems}, we should pay special attention when we calculate the terms  
 for which $\alpha=\beta$. For the 1D dimensional case, it is proposed to estimate the average value from the closest neighbors \cite{Zhao2018ConstructionProblems}:
\begin{multline} \label{eq:PD_discret_nei_1} 
\frac{1}{2} \left[\Lambda \left(\textbf{x}_\alpha ,\textbf{x}_{\alpha+1} ,t\right)\frac{T\left(\textbf{x}_{\alpha+1} ,t\right)-T\left(\textbf{x}_\alpha ,t\right)}{\left\| \textbf{x}_{\alpha+1} -\textbf{x}_\alpha \right\| ^{2} } \right. +\\
\Lambda \left(\textbf{x}_\alpha ,\textbf{x}_{i-1} ,t\right)\frac{T\left(\textbf{x}_{i-1} ,t\right)-T\left(\textbf{x}_\alpha ,t\right)}{\left\| \textbf{x}_{i-1} -\textbf{x}_\alpha \right\| ^{2} } -
\\
\rho_w C_w \left(\textbf{x}_\alpha ,t\right) U \left(\textbf{x}_\alpha ,\textbf{x}_{i+1} ,t\right)\frac{T\left(\textbf{x}_{i+1} ,t\right)-T\left(\textbf{x}_\alpha ,t\right)}{\left\| \textbf{x}_{i+1} -\textbf{x}_\alpha \right\| } - \\
\left. \rho_w C_w \left(\textbf{x}_\alpha ,t\right) U \left(\textbf{x}_\alpha ,\textbf{x}_{i-1} ,t\right)\frac{T\left(\textbf{x}_{i-1} ,t\right)-T\left(\textbf{x}_\alpha ,t\right)}{\left\| \textbf{x}_{i-1} -\textbf{x}_\alpha \right\| } \right]
\end{multline} 
For the 2D case, we consider the eight neighbor nodes of the node $\textbf{x}_\alpha$:
\begin{multline}
\label{eq:PD_discret_nei_2} 
\frac{1}{8} \sum _{\beta=1}^{8}\Lambda \left(\textbf{x}_\alpha ,\textbf{x}_\beta ,t\right)\frac{T\left(\textbf{x}_\beta ,t\right)-T\left(\textbf{x}_\alpha ,t\right)}{\left\| \textbf{x}_\beta -\textbf{x}_\alpha \right\| ^{2} }  - \\
\frac{1}{8} \rho_w C_w \sum _{\beta=1}^{8} U \left(\textbf{x}_\alpha ,\textbf{x}_\beta ,t\right)\frac{T\left(\textbf{x}_\beta ,t\right)-T\left(\textbf{x}_\alpha ,t\right)}{\left\| \textbf{x}_\beta -\textbf{x}_\alpha \right\| }
\end{multline}

\section{Verification}\label{sec:verify}
To assess the accuracy of the implementation, we consider several problems, where we calculate the temperature distribution at different time instances and compare the results with alternative solutions obtained using analytical expressions and/or the FEM. The first problem was designed to verify the implementation of heat conduction with phase change only; the second problem was chosen to test the implementation of convective heat transport only; and the third problem was included to test the implementation of fully coupled convective-conductive heat transport with phase change under pressure driven water flow. In addition we analyse a fourth problem, involving high water velocities, to demonstrate the application of our implementation for situations that are challenging for numerical methods that are based on local (differential) formulations.

In all problems, the medium undergoing phase change is water. The thermo-physical properties of the three phases - solid, water and ice - are presented in Table \ref{table:thermo}, where data was taken from \cite{Grenier2018GroundwaterCases}. The PD boundary conditions were applied using a boundary layer with thickness equal to the horizon radius $\delta$ that was added at the boundary ends in 1D problems, and along the domain perimeter in 2D problem (details are given in \cite{Zhao2018ConstructionProblems}). 

\begin{table}
\caption{Thermo-physical properties of phases in test problems}
\centering
\begin{tabular}{c c} \hline
Physical properties & Parameter values \\ \hline 
Porosity, n & 0.3 \\ 
Thermal conductivity of liquid water, $\lambda_w $ , (W m${}^{-1}$ $^\circ {\rm C}$${}^{-1}$) & 0.6 \\ 
Thermal conductivity of ice, $\lambda_i $ , (W m${}^{-1}$ $^\circ {\rm C}$${}^{-1}$) & 2.14 \\ 
Thermal conductivity of solid grains, $\lambda_s $ , (W m${}^{-1}$ $^\circ {\rm C}$${}^{-1}$) & 9 \\ 
Specific heat capacity of liquid water, $C_w $ , (J kg${}^{-1}$ $^\circ {\rm C}$${}^{-1}$) & 4182 \\ 
Specific heat capacity of ice, $C_i $ , (J kg${}^{-1}$ $^\circ {\rm C}$${}^{-1}$) & 2060 \\ 
Specific heat capacity of solid grains, $C_s $ , (J kg${}^{-1}$ $^\circ {\rm C}$${}^{-1}$) & 835 \\ 
Liquid water density, $\rho_w $ , (kg m${}^{-3}$) & 1000 \\
Ice density, $\rho_i $ , (kg m${}^{-3}$) & 920 \\ 
Solid grain density, $\rho_s $ , (kg m${}^{-3}$) & 2650 \\ 
Dynamic viscosity of liquid water, $\mu $ , (kg m${}^{-1}$ s${}^{-1}$) & 1.793$\mathrm{\cdot}$10${}^{-3}$ \\ 
Latent heat of solidification, $L$ , (J kg${}^{-1}$) & 334000 \\ 
Residual saturation, $S_{w_{res} } $  & 0 \\ 
$\Delta T$ , $^\circ {\rm C}$ & 0.5 \\ 
Water solidification temperature, $T_f $ , $^\circ {\rm C}$ & 0.0 \\
Intrinsic permeability, $k_{int} $ , m${}^{2}$ & 1.3$\mathrm{\cdot}$10${}^{-}$${}^{10}$ \\
$\Omega $  & 50 \\ 
\end{tabular}
\label{table:thermo}
\end{table}

\subsection{Heat transfer with phase change: 1D problem}\label{sec:verify_diff_phase}
The problem of heat transfer with phase change has an analytical solution in 1D. The problem is formulated for a semi-infinite domain, $0 \leq x \leq \infty$, with initial condition $T(x,0)=T_{\infty}$ and boundary conditions $T(0,t)=T_0$ (constant source temperature) and $T(\infty,t)=T_{\infty}$. The solution is \cite{Hahn2012HeatConduction}:

\begin{equation} \label{eq:Class_solution} 
\begin{array}{l} {T\left(x,t\right)=\left\{\begin{array}{c} {T_0+\left(T\left(x_c,t\right)-T_0\right)\frac{{\rm erf}\left(\frac{x}{\left(4\alpha_{i} t\right)^{1/2} } \right)}{{\rm erf}\left(\frac{\beta }{\left(4\alpha_{i} \right)^{1/2} } \right)} ,\quad 0\le x\le x_c \left(t\right)} \\ {T_{\infty}+\left(T\left(x_c ,t\right)-T_{\infty}\right)\frac{{\rm erfc}\left(\frac{x}{\left(4\alpha _{w} t\right)^{1/2} } \right)}{{\rm erfc}\left(\frac{\beta }{\left(4\alpha _{w} \right)^{1/2} } \right)} ,\quad x_c \left(t\right)\le x\le \infty } \end{array}\right. } \\ {} \end{array} 
\end{equation} 
where $x_c \left(t\right)=\beta \sqrt{\alpha _{i} t} $ is the position of the freezing front at time $t$; $\alpha_{i} =\lambda_i /\left(C_i \rho_w \right)$ and $\alpha_{w} =\lambda_w /\left(C_w \rho_w \right)$ are the thermal conductivities of ice and water, respectively; ${\rm erf}(x)=\frac{2}{\sqrt{\pi } } \int _{0}^{x}e^{-p^{2} } dp $ is the error function; ${\rm erfc}(x)=1-{\rm erf}(x)$ is the complementary error function; and $\beta$ is the root of the transcendental equation:
\begin{multline} \label{eq:Transc_eq} 
\frac{e^{-\beta ^{2} } }{{\rm erf}\left(\beta \right)} +\frac{\lambda_w }{\lambda_i } \left(\frac{\alpha _{i} }{\alpha _{w} } \right)^{1/2} \frac{T\left(x_c ,t\right)-T\left(\infty ,t\right)}{T\left(x_c ,t\right)-T\left(0,t\right)} \frac{e^{-\beta ^{2} \left(\alpha _{i} /\alpha _{w} \right)} }{{\rm erfc}\left(\beta \sqrt{\alpha _{i} /\alpha _{w} } \right)} = \\
\frac{\beta L\sqrt{\pi } }{C_i \left[T\left(x_c ,t\right)-T\left(0,t\right)\right]}  
\end{multline} 

We considered the freezing of water and compared our result with the solution provided by Eq. (\ref{eq:Class_solution}). The simulated domain had finite length of 0.3 m. The particle size was $\Delta x= 0.002$ m, the horizon radius was $\delta =0.006$ m, and the time step was $\Delta t=2$ s. In the calculations we neglected the change of density during the phase change, since the analytical solution did not take it into account. Hence, the density of liquid and solid water was taken as $\rho_w$. The specific values of the initial and boundary conditions were $T\left(x,0\right) = T\left(\infty ,t\right)= 8 ^\circ {\rm C}$ and $T\left(0,t\right) = -20{}^\circ {\rm C}$. Equation (\ref{eq:PD_mass_space_disc}) was solved with $n = 1$ (no solid phase) and $u \left(\mathbf{x}_{\beta} ,\mathbf{x}_{\alpha},t \right)=0$ (no water flow). This model simulated a physical time for the process of 10 hours.

\begin{figure}
     \centering
     \begin{subfigure}[b]{0.7\textwidth}
         \centering
         \includegraphics[width=1\textwidth]{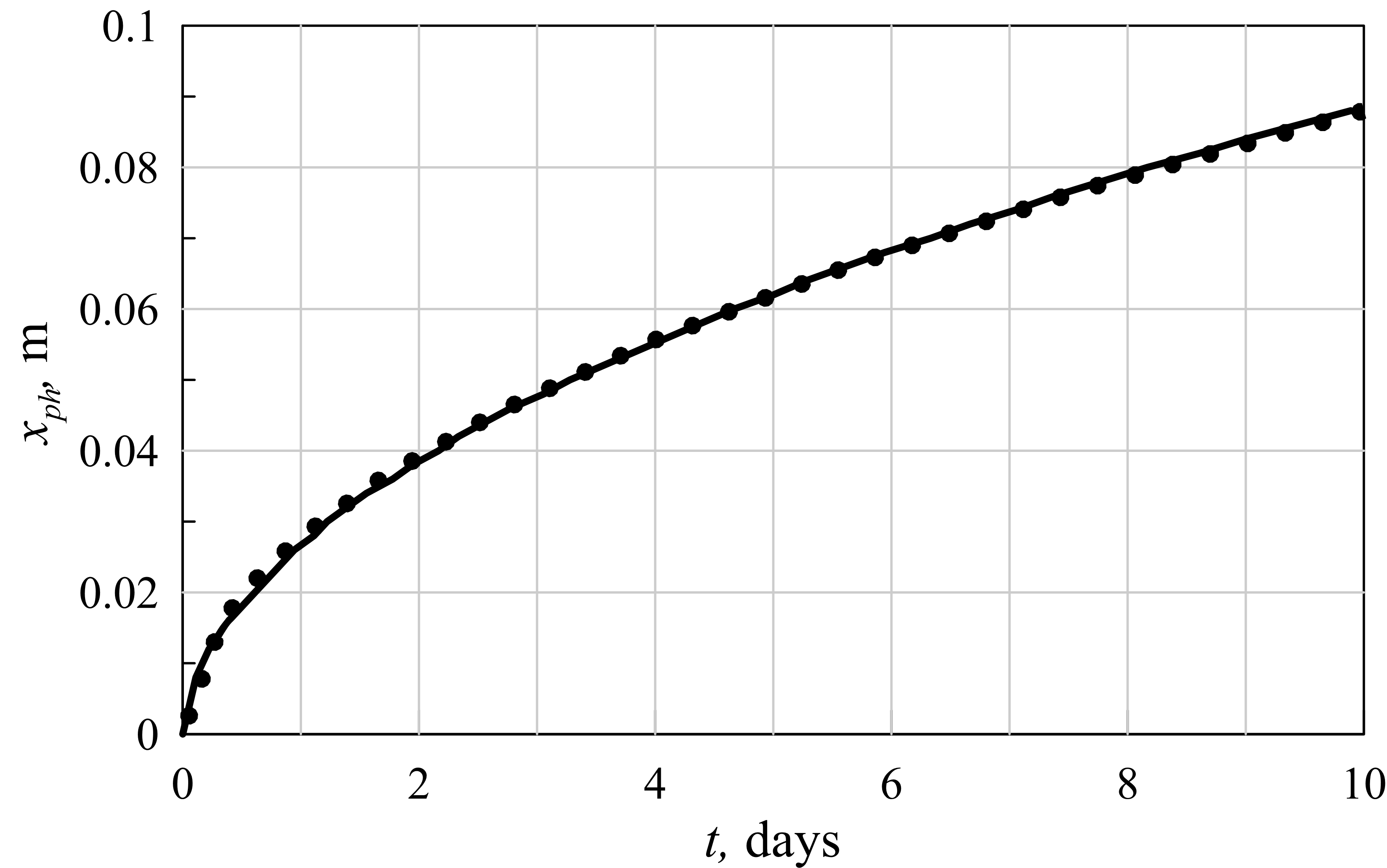}
         \label{fig:y equals x}
     \end{subfigure}
     \vfill
     \begin{subfigure}[b]{0.7\textwidth}
         \centering
         \includegraphics[width=1\textwidth]{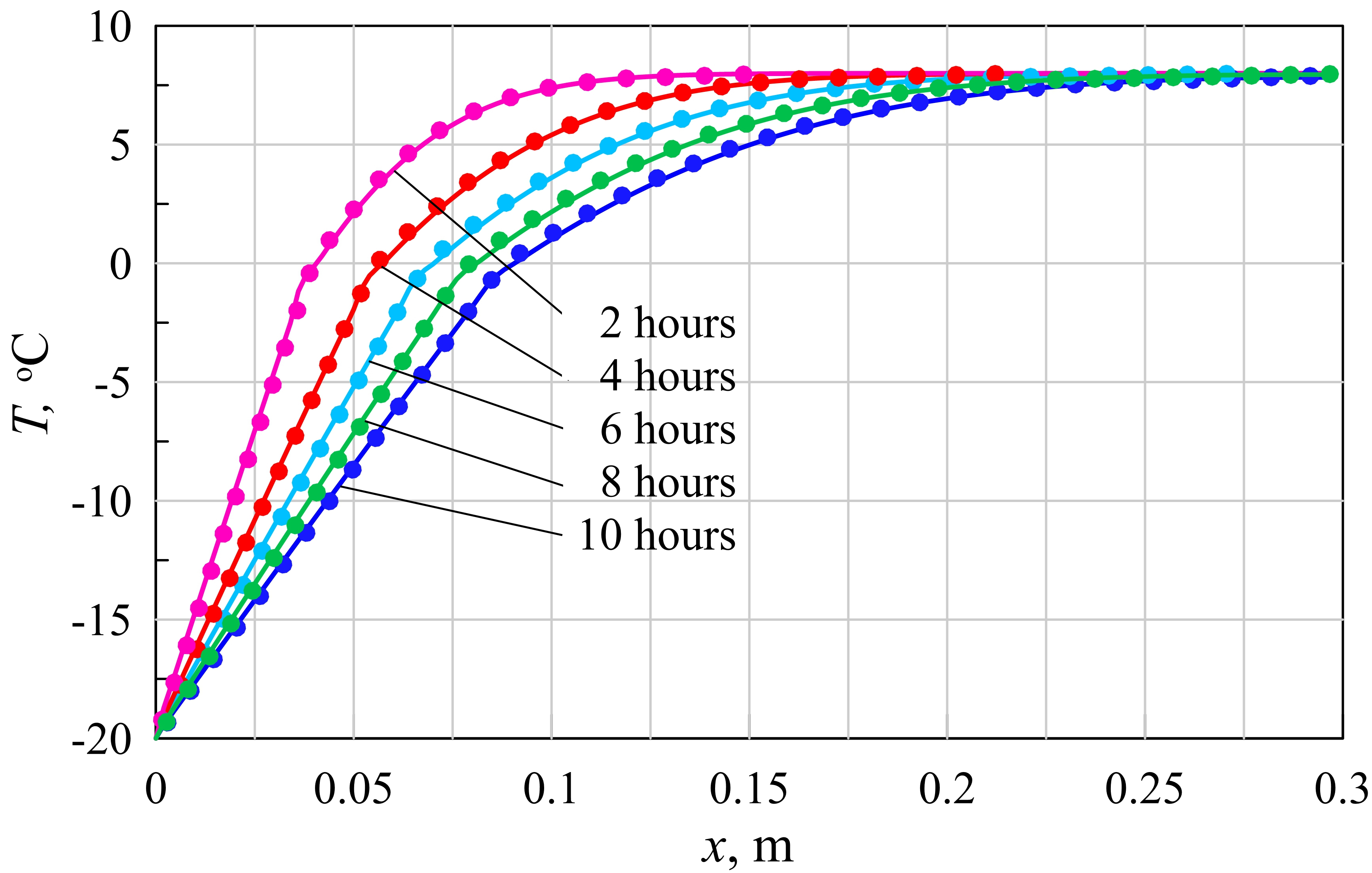}
         \label{fig:three sin x}
     \end{subfigure}
        \caption{One-dimensional heat transfer with phase change according to the PD model and analytical results. The upper figure shows the phase front propagation. The lower figure shows the profiles of temperature distribution for different times. The solid lines are the developed PD solution; the dots are the analytical solution.}
        \label{fig:PD_heat_transfer_ph_ch}
\end{figure}

It is noted that in this problem we compare the temperature field in an infinite domain, given by the analytical solution, with the temperature field in a finite domain, provided by the PD solution. The results are presented in Fig. \ref{fig:PD_heat_transfer_ph_ch}, where the left graph shows the displacement of the phase front with time, and the right graph shows the temperature distributions at time instances 2, 4, 6, 8 and 10 hours. There is very close agreement between the two solutions, providing further confidence on the implementation of heat conduction with phase change in the bond-based PD model. It is noted that the proposed bond-based solution provides results with the same accuracy as the PDDO model that was developed in \cite{Madenci2017}.

\subsection{Heat transfer with water flow: 1D problem}\label{sec:verify_conv_diff}
This problem of convective-conduction thermal diffusion also has an analytical solution in 1D. The problem is again formulated for a semi-infinite domain, $0 \leq x \leq \infty$, with initial condition $T(x,0)=T_{\infty}$ and boundary conditions $T(0,t)=T_0$ (constant source temperature) and $T(\infty,t)=T_{\infty}$, supplemented by fluid velocity, $v$. The solution is given by \cite{Hahn2012HeatConduction}:
\begin{equation} \label{eq:an_water_flow)} 
T\left(x,t\right)-T_{\infty}=\frac{T_0}{2} \left({\rm erfc}\left(\frac{x-vt}{2\sqrt{\alpha _{w} t} } \right)+\exp \left(\frac{vt}{\alpha _{w} } \right){\rm erfc}\left(\frac{x+vt}{2\sqrt{\alpha _{w} t} } \right)\right) 
\end{equation} 

We simulated the process in a domain of finite length 0.4 m, with particle size $\Delta x= 0.005$ m, horizon radius $\delta =0.015$ m, and time step $\Delta t=10$ s. The specific values of the initial and boundary conditions were $T\left(x,0\right) = T\left(\infty ,t\right)= 40^{\circ} {\rm C}$ and $T\left(0,t\right) = 5^{\circ} {\rm C}$. Furthermore, the water was assumed to have constant velocity $v=0.0025 $ cm/s.

\begin{figure}[ht]
    \centering
    \includegraphics[width=0.7\textwidth]{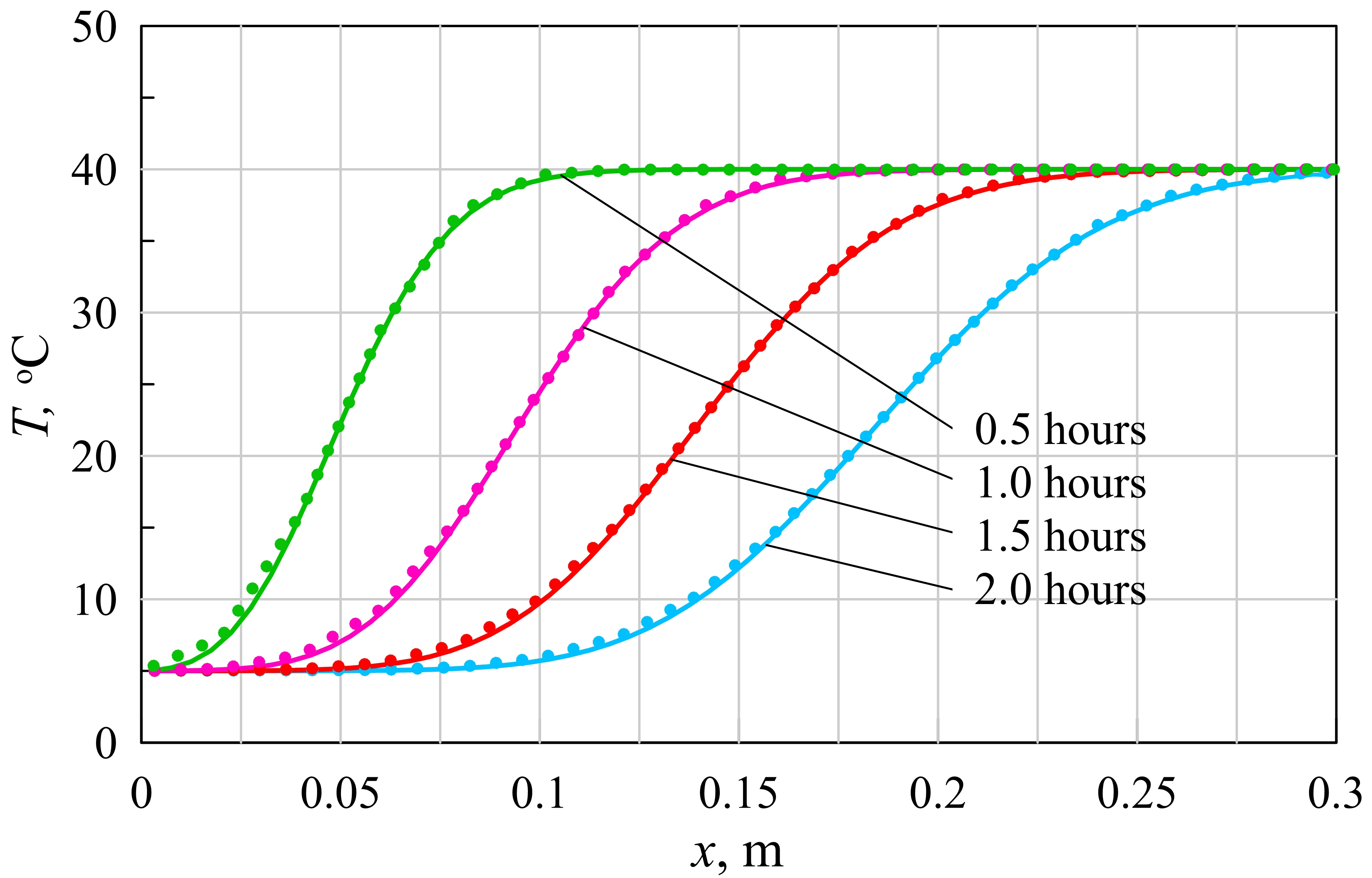}
    \caption{One-dimensional heat transfer with water flow according to the PD model and analytical results. The figure shows the profiles of temperature distribution for different time intervals. The solid lines are the developed PD solution; the dots are the analytical solution.}
    \label{fig:PD_heat_transfer_water_flow}
\end{figure}

The results from our simulations and from the analytical solution are presented in Fig.  \ref{fig:PD_heat_transfer_water_flow} for time instances 0.5, 1, 1.5 and 2 hours. There is close agreement between the two alternative solutions which provides further confidence that the PD formulation for convective-conductive heat transport has been correctly implemented. These results are also in close  agreement with similar solutions for chemical diffusion-advection problems presented in \cite{Zhao2018ConstructionProblems, Yan2020PeridynamicsMedia} . 

\subsection{Heat transfer with water flow and phase change: 2D problem}\label{sec:verify_full}
In this sub-section we assess the performance of the developed model applied to heat transfer with phase change in a fully saturated porous medium with pressure driven water flow. Analytical solutions for such problems do not exist and we compare our PD results, with results obtained using the finite element method (Comsol Multiphysics \cite{Grenier2018GroundwaterCases}).

\begin{figure}[ht]
    \centering
    \includegraphics[width=0.7\textwidth]{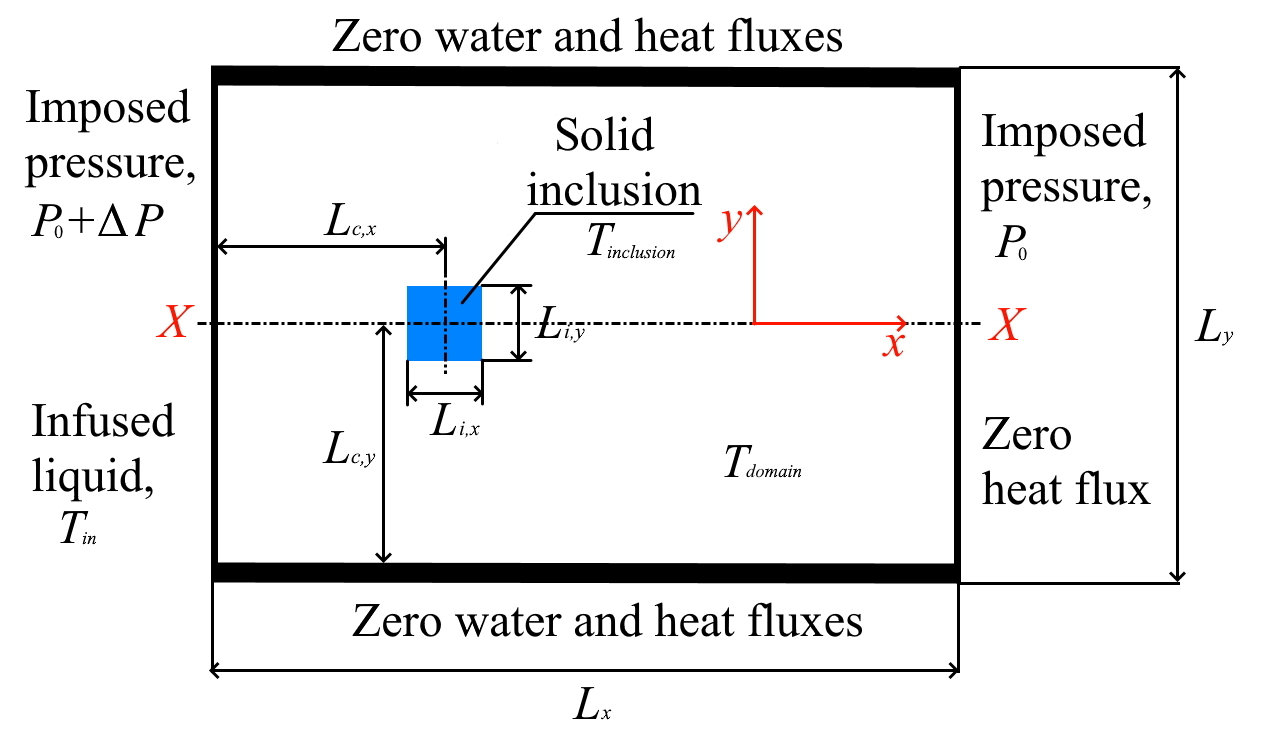}
    \caption{Schematic of frozen inclusion problem: geometry, boundary and initial conditions}
    \label{fig:PD_heat_2D}
\end{figure}

We adopted a benchmark problem proposed in \cite{Grenier2018GroundwaterCases} to estimate the efficacy and accuracy of different FEM packages. This problem describes the thawing of an initially frozen inclusion in a porous medium, subject to pressure driven water flow under constant positive temperature. The domain and its boundary conditions are illustrated in Fig. \ref{fig:PD_heat_2D} and geometric and physical parameters are presented in Table \ref{table:inclusion}. We note that these values are selected for our work and differ from the test problem in \cite{Grenier2018GroundwaterCases}.

\begin{table}
\caption{Parameters of the frozen inclusion model}
\centering
\begin{tabular}{c c} \hline 
Parameter & Value  \\ \hline 
Domain length, $L_x$ , m & 0.75  \\ 
Domain height, $L_y$ , m & 0.5 \\ 
Horizontal position of inclusion centre, $L_{cx} $ , m & 0.28 \\ 
Vertical position of inclusion centre, $L_{cy} $ , & 0.25 \\ 
Inclusion length, $L_{i,x} $  & 0.075 \\ 
Inclusion height, $L_{i,y} $  & 0.075 \\ 
Initial temperature of unfrozen domain, $T_{domain} $ , $^\circ {\rm C}$ & 15.0 \\ 
Initial temperature of frozen inclusion, $T_{inclusion} $ , $^\circ {\rm C}$ & -5.0 \\ 
Temperature of infused liquid, $T_{in} $ , $^\circ {\rm C}$ & 15.0 \\ \
Imposed pressure (3th / 4th example) $P_0 $ ,Pa & 5 / 150 \\ 
Excess pressure (3th / 4th example), $\Delta P$ , Pa & 370 / 11100 \\ 
\end{tabular}
\label{table:inclusion}
\end{table}

In our simulations, the particle size was $\Delta x = 0.005$ m, the horizon radius was $\delta = 0.015$ and the time step was $\Delta t=5$ s. In this problem, we assumed that the t-bonds in the model can be divided into groups: impermeable solid, liquid and interfacial; for more details, see Section 3. A t-bond was classified as impermeable solid if it connects nodes with temperature $T \leq T_f - \Delta T$. We supposed that $\Delta T = 0.5 ^\circ {\rm C}$ by analysing the soil freezing characteristic curve (\ref{eq:water_saturation}) and the relation for relative permeability (\ref{eq:rel_perm}). It showed that by this temperature the relative permeability of soils  $k_{rel}$ became negligible. Therefore, these impermeable bonds were not used for fluid transfer, and the nodes with such temperature were not considered in pressure estimation.

\begin{figure}[ht]
 
    \centering
     \begin{subfigure}[b]{0.49\textwidth}
         \centering
         \includegraphics[width=0.85\textwidth]{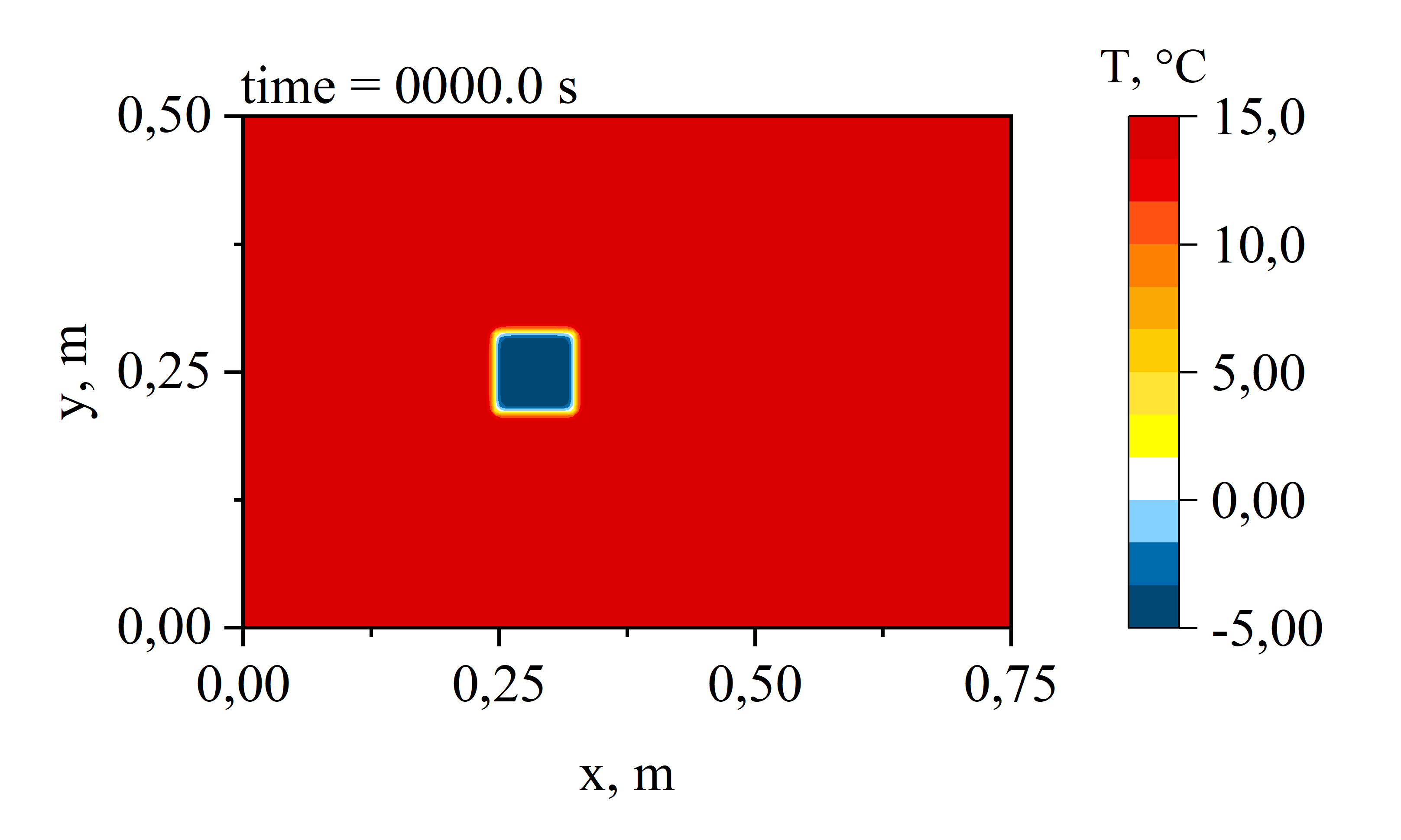}
     \end{subfigure}
     \hfill
     \begin{subfigure}[b]{0.49\textwidth}
         \centering
         \includegraphics[width=0.85\textwidth]{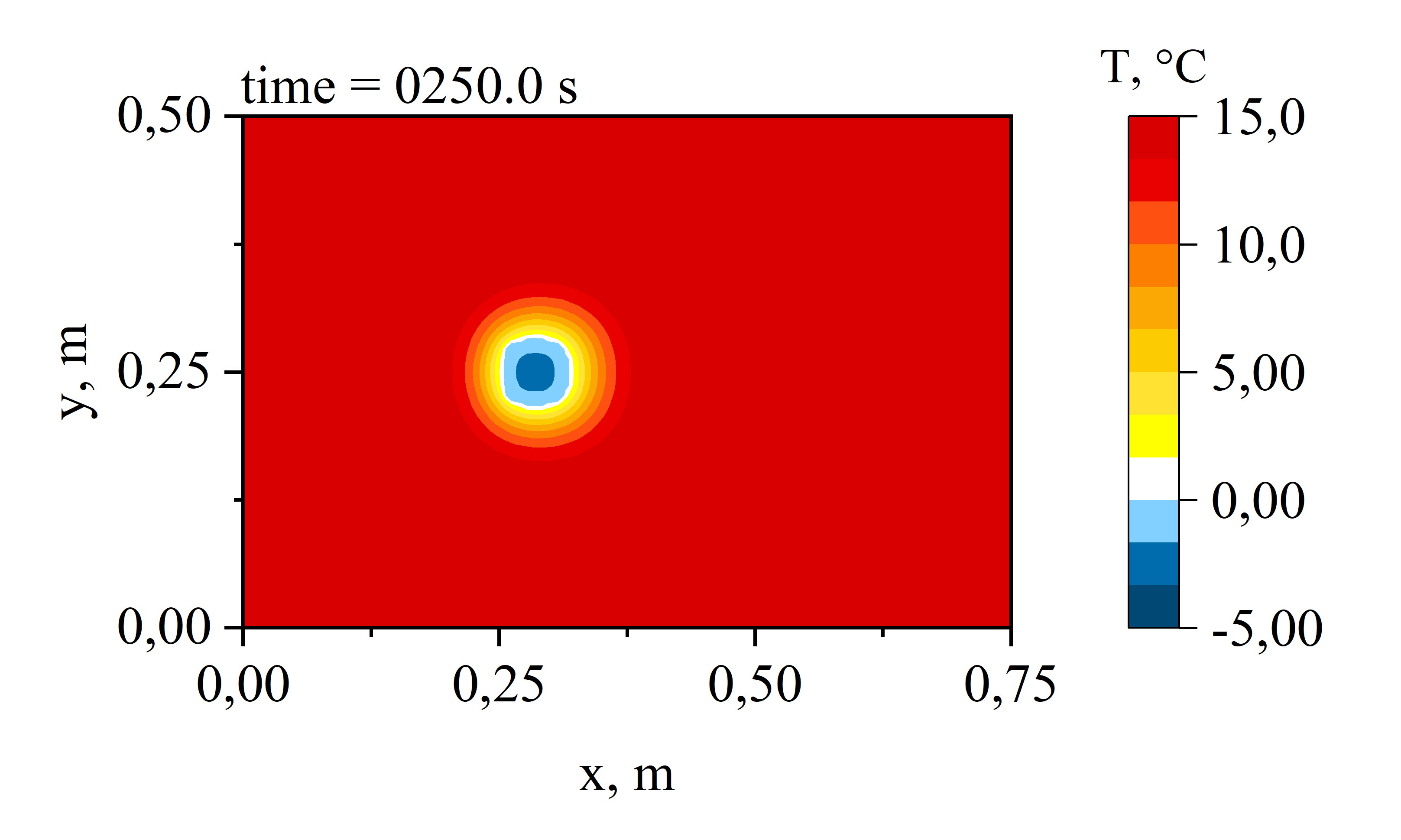}
     \end{subfigure}
     \vfill
    \begin{subfigure}[b]{0.49\textwidth}
         \centering
         \includegraphics[width=0.85\textwidth]{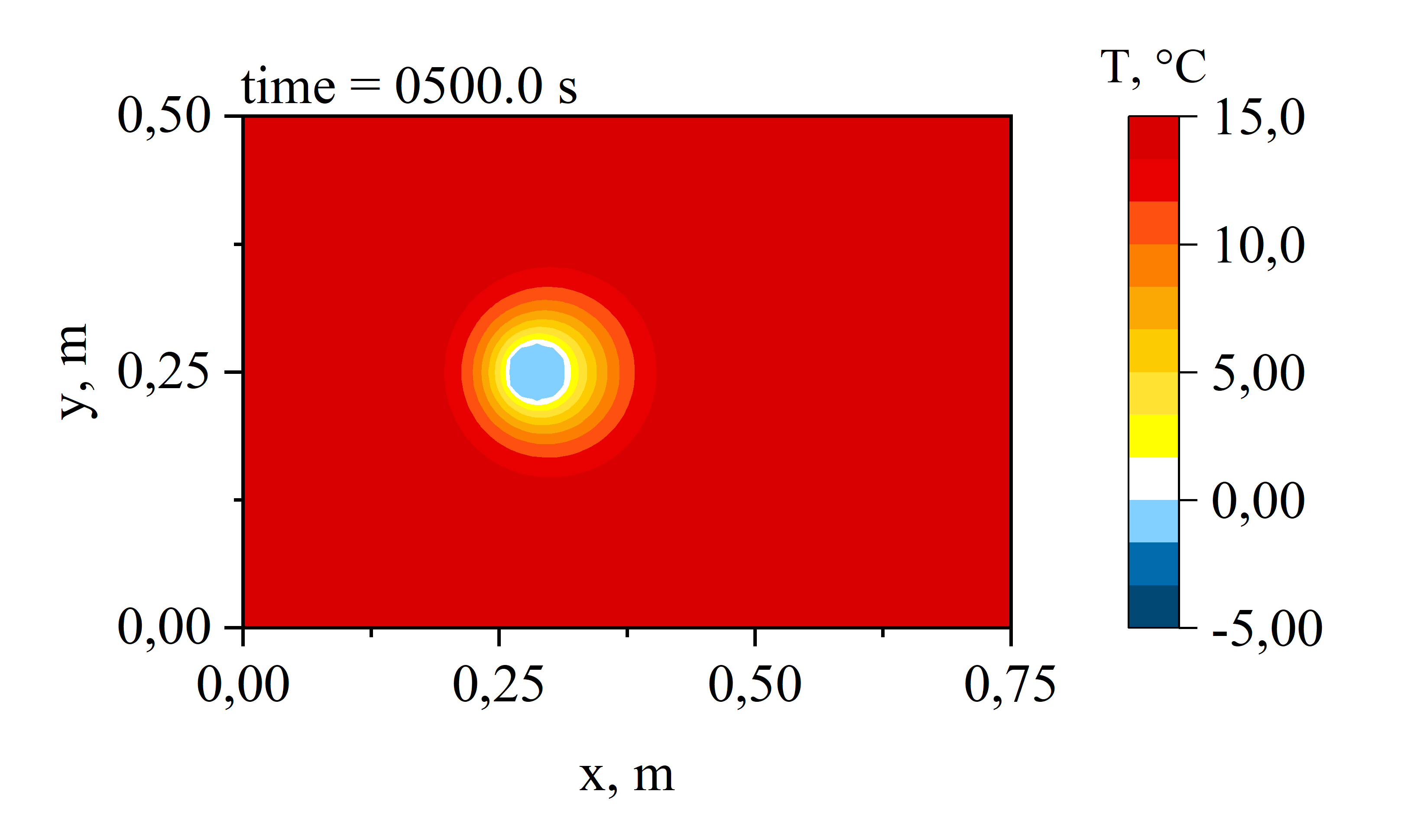}
     \end{subfigure}
     \hfill
     \begin{subfigure}[b]{0.49\textwidth}
         \centering
         \includegraphics[width=0.85\textwidth]{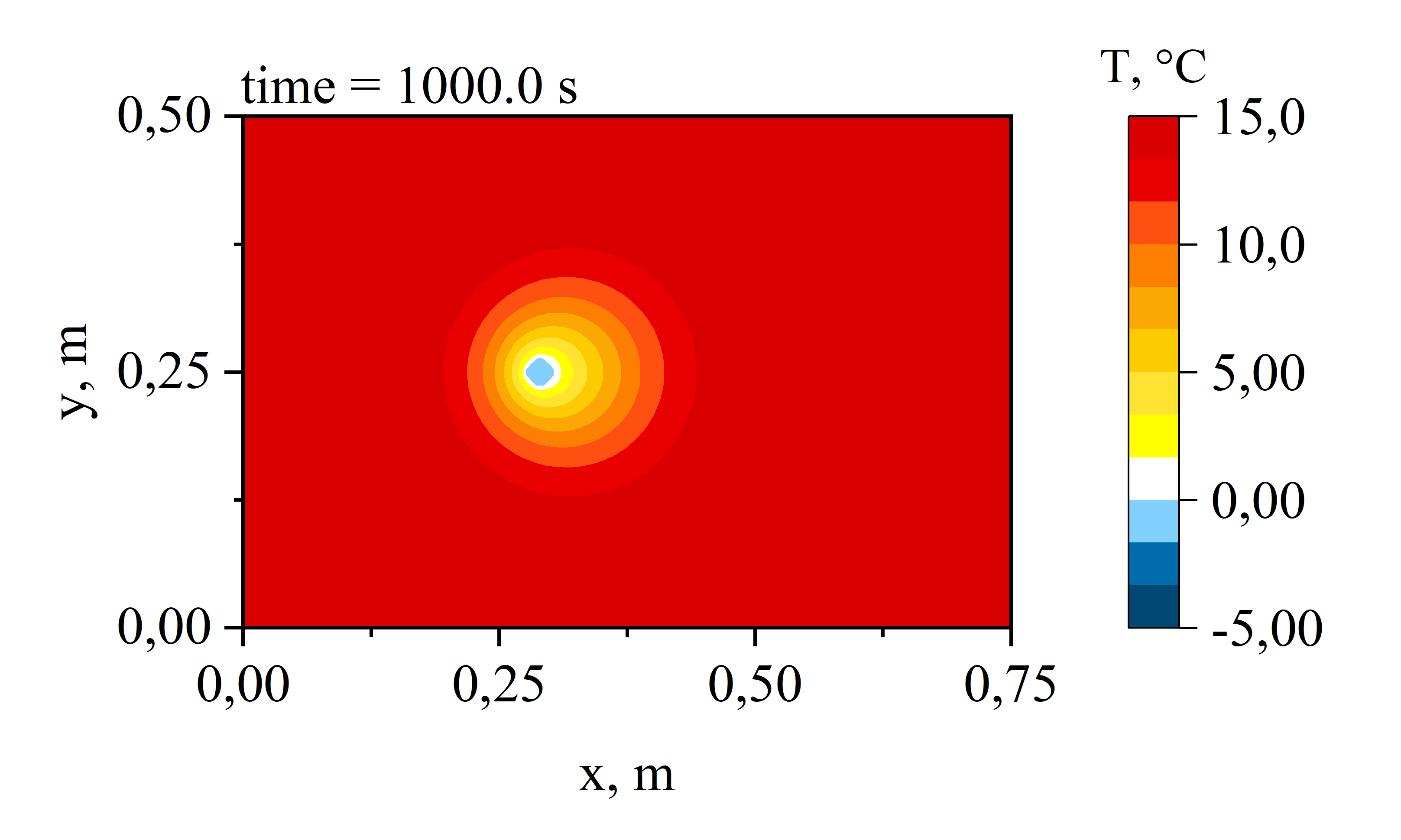}
     \end{subfigure}
          \vfill
    \begin{subfigure}[b]{0.49\textwidth}
         \centering
         \includegraphics[width=0.85\textwidth]{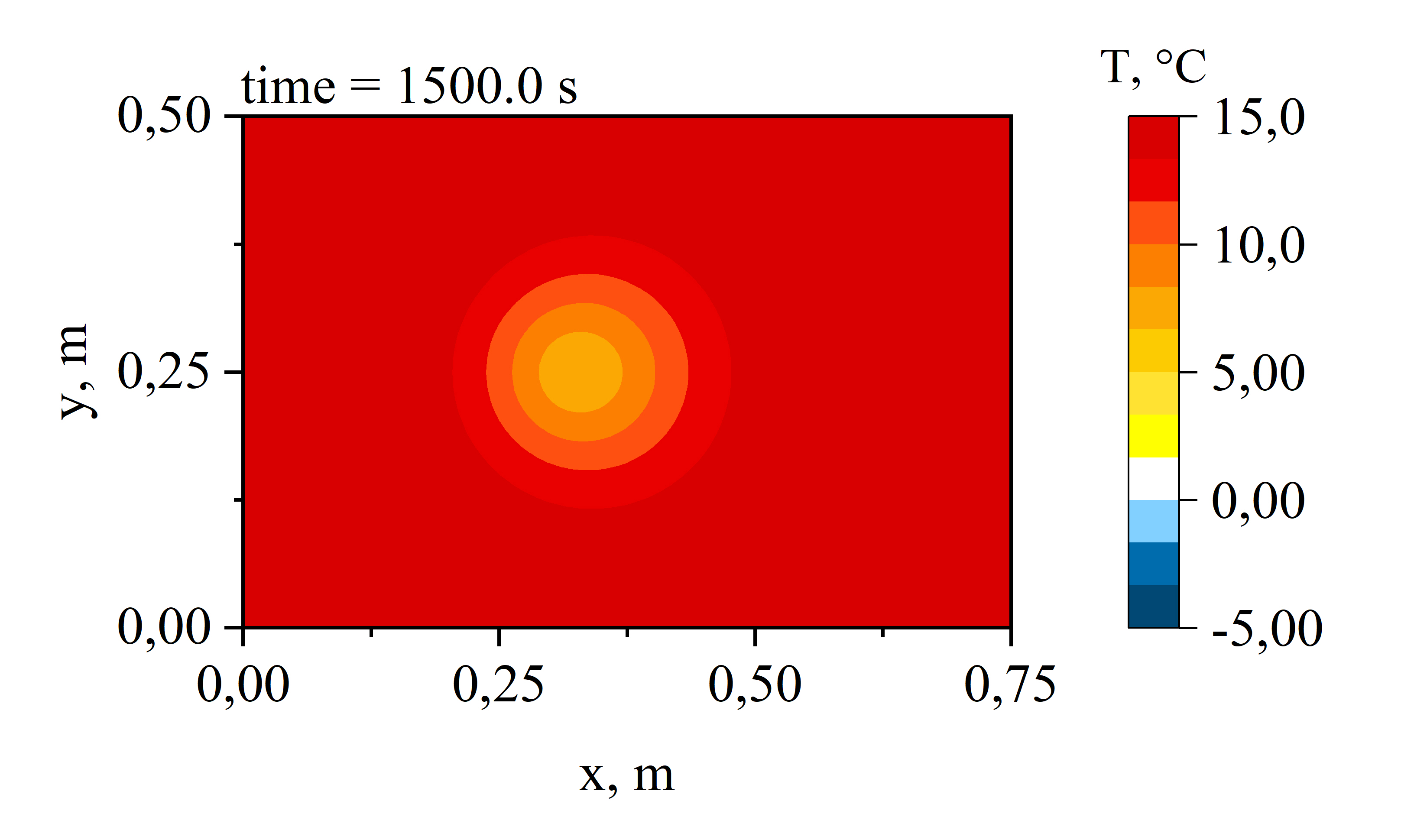}
     \end{subfigure}
     \hfill
     \begin{subfigure}[b]{0.49\textwidth}
         \centering
         \includegraphics[width=0.85\textwidth]{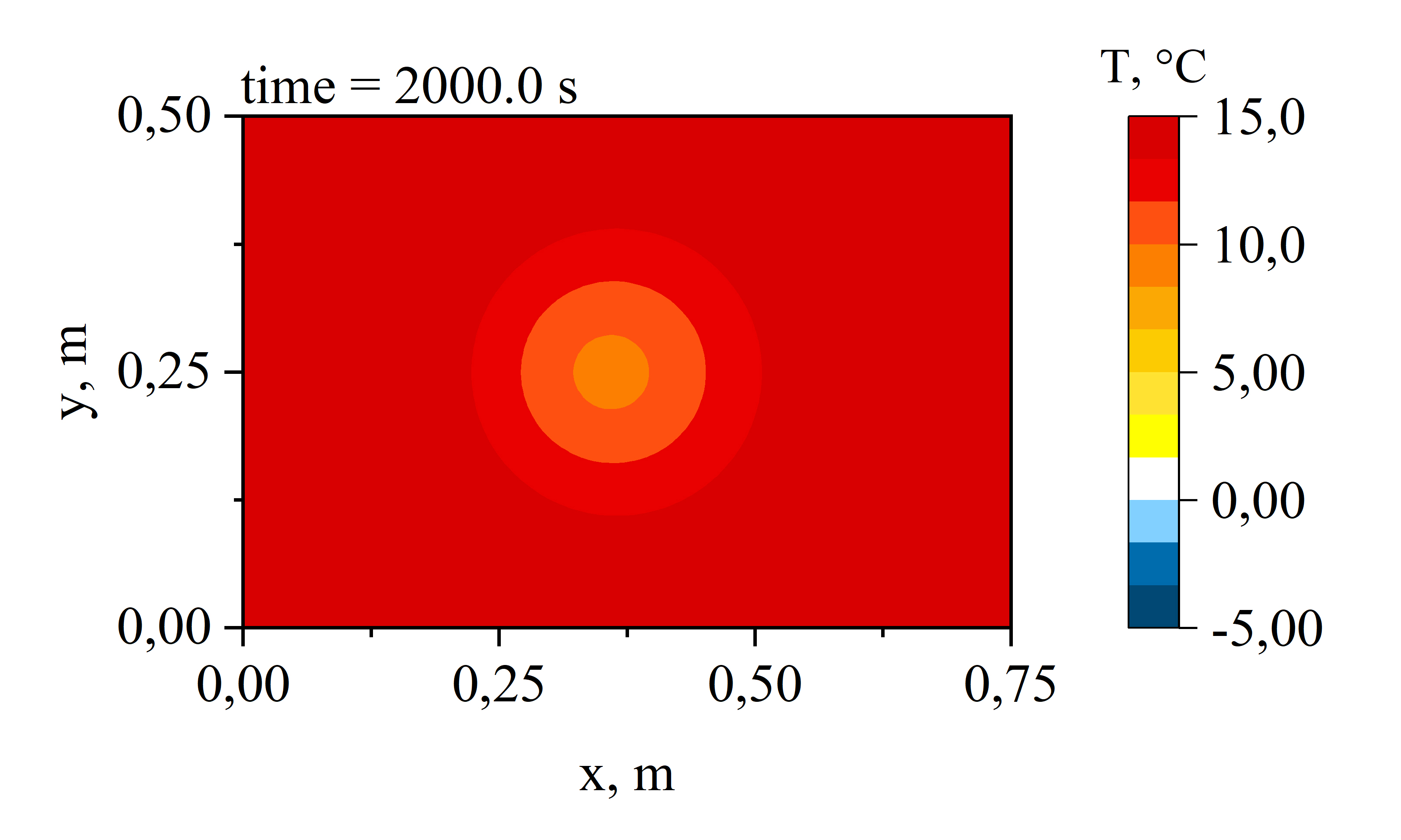}
     \end{subfigure}
    \vfill
    \begin{subfigure}[b]{0.49\textwidth}
         \centering
         \includegraphics[width=0.85\textwidth]{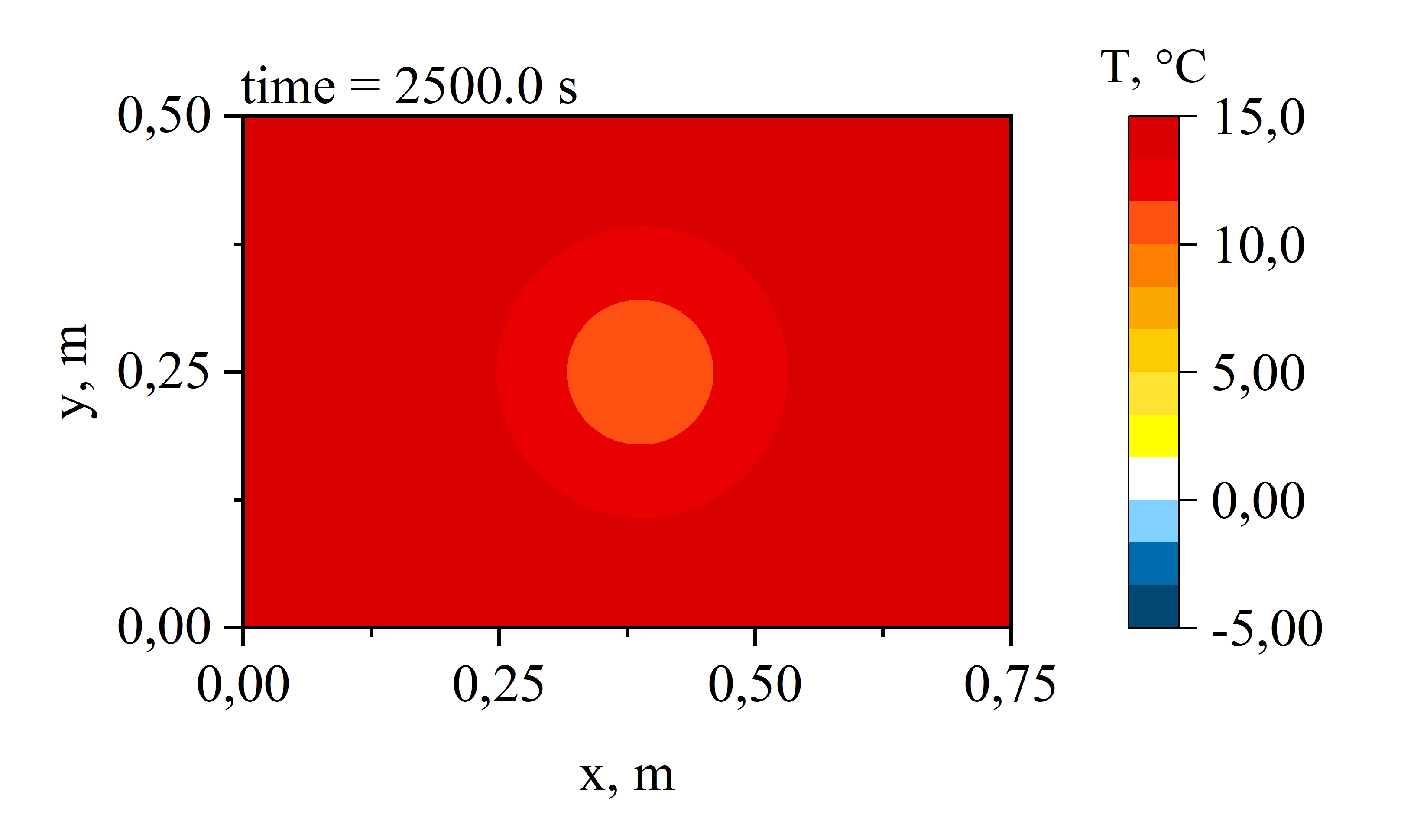}
     \end{subfigure}
     \hfill
     \begin{subfigure}[b]{0.49\textwidth}
         \centering
         \includegraphics[width=0.85\textwidth]{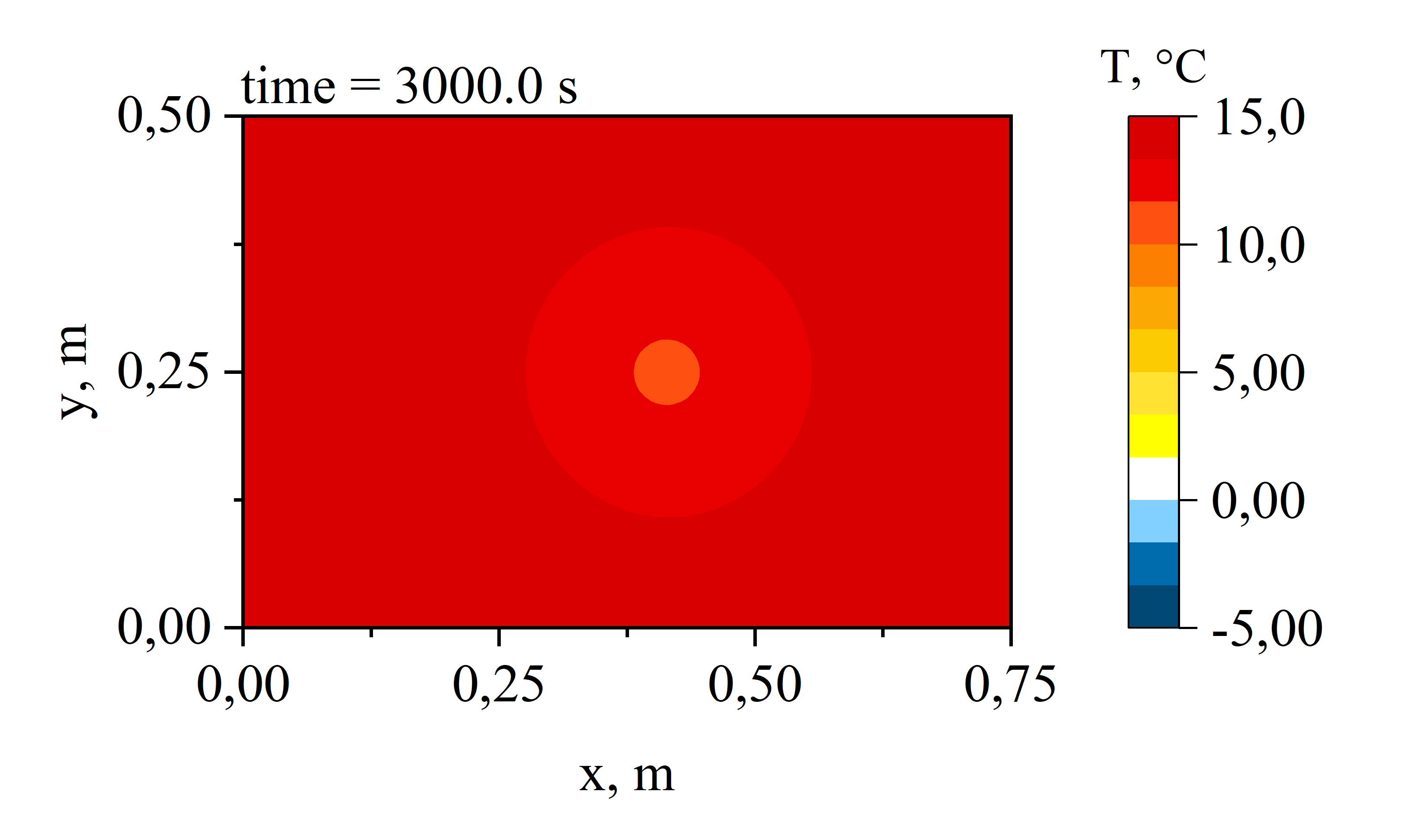}
     \end{subfigure}
    \caption{Temperature distribution, $^\circ {\rm C}$, at different moments of time calculated with the proposed PD model}
    \label{fig:PD_heat_2D_temp_distr}
\end{figure}

\begin{figure}[ht]
    \centering
    \includegraphics[width=0.7\textwidth]{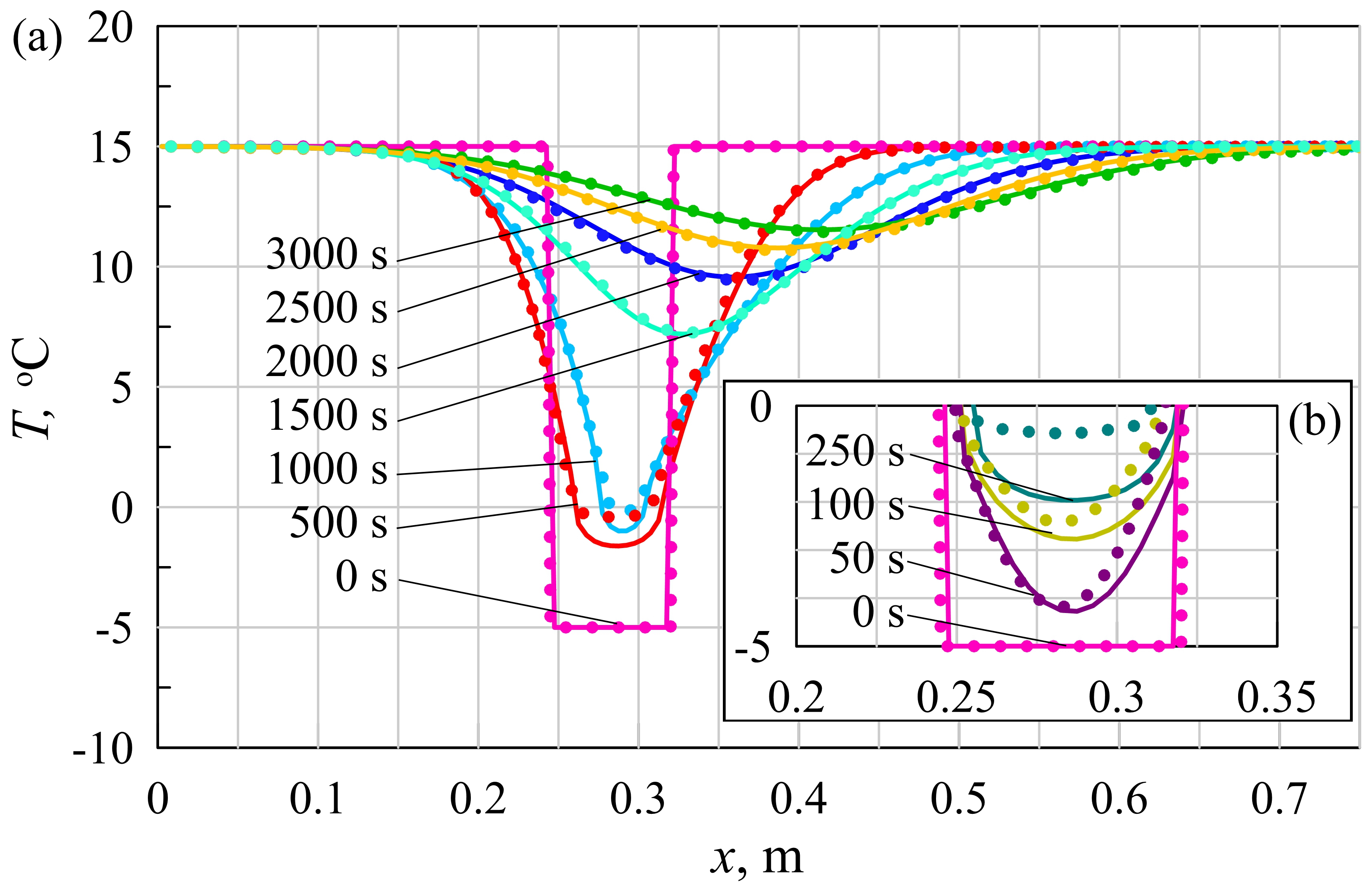}
    \caption{Temperature distribution along the X--X cross section at different moments of time, calculated by PD and FEM simulations. The solid lines are the developed PD solution; the dots are the analytical solution. (a) is the temperature distribution within the domain; (b) is the temperature distribution within the inclusion.}
    \label{fig:PD_heat_2D_temp_distr_x}
\end{figure}

\begin{figure}[ht]
    \centering
     \begin{subfigure}[b]{0.49\textwidth}
         \centering
         \includegraphics[width=0.85\textwidth]{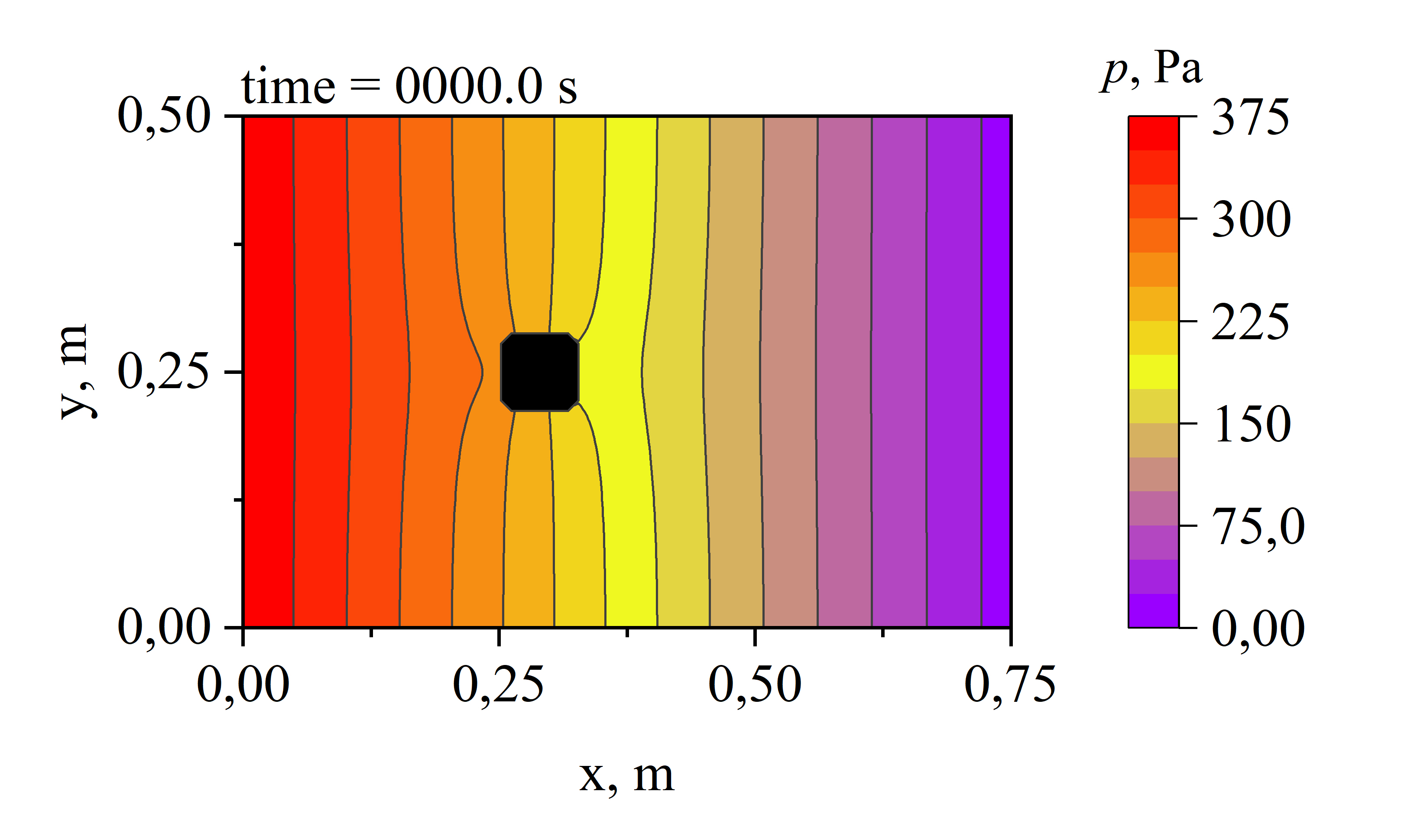}
     \end{subfigure}
     \hfill
     \begin{subfigure}[b]{0.49\textwidth}
         \centering
         \includegraphics[width=0.85\textwidth]{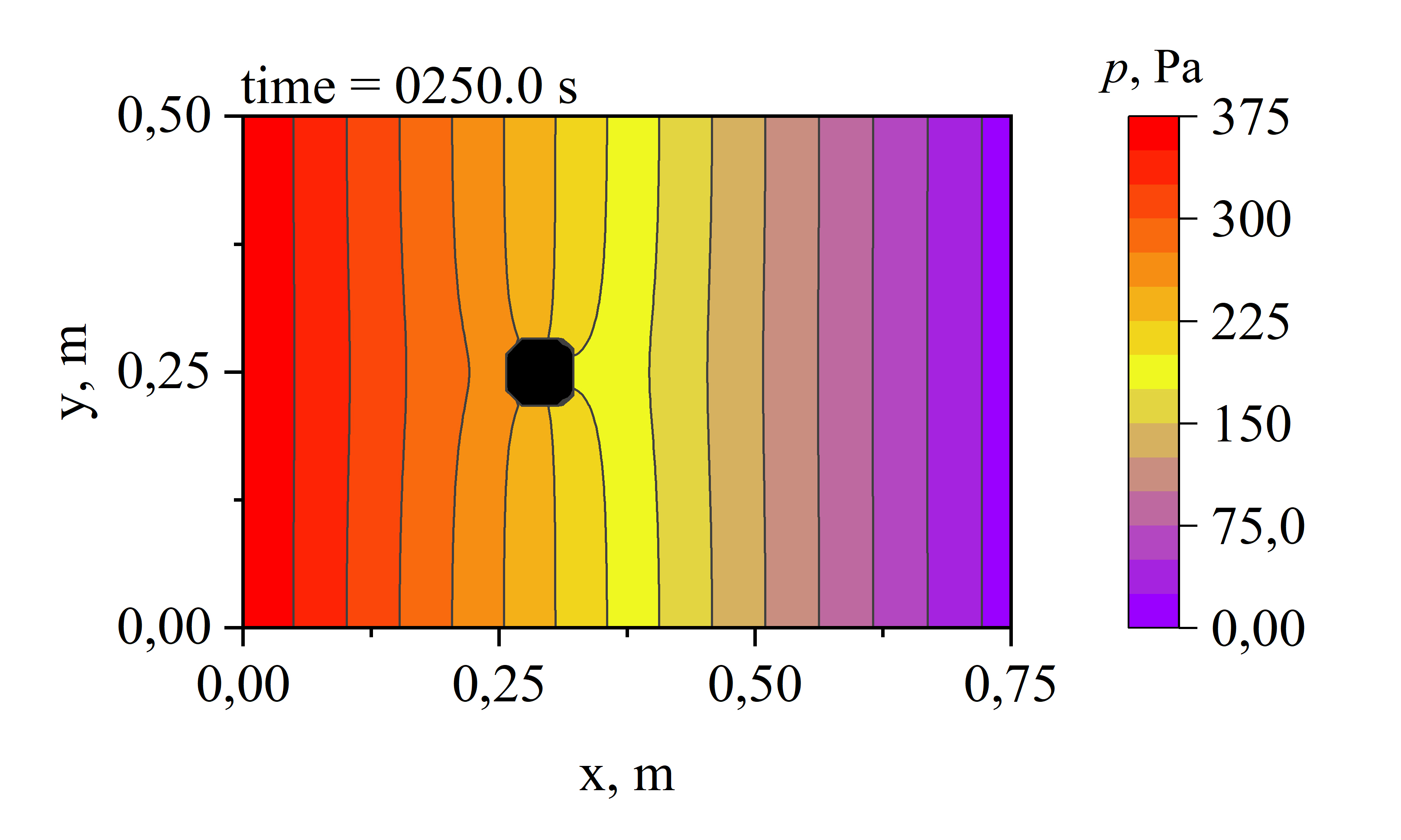}
     \end{subfigure}
     \vfill
    \begin{subfigure}[b]{0.49\textwidth}
         \centering
         \includegraphics[width=0.85\textwidth]{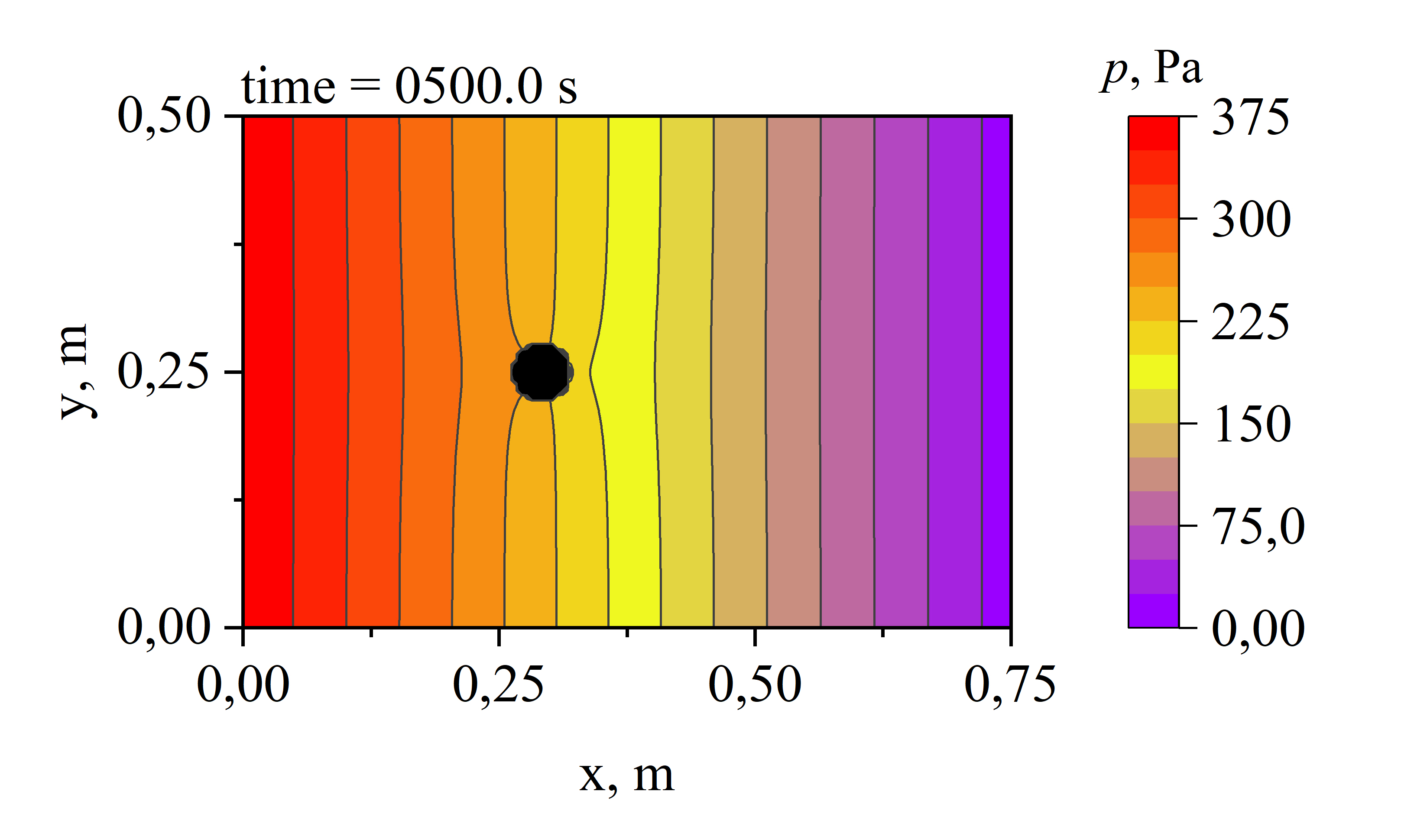}
     \end{subfigure}
     \hfill
     \begin{subfigure}[b]{0.49\textwidth}
         \centering
         \includegraphics[width=0.85\textwidth]{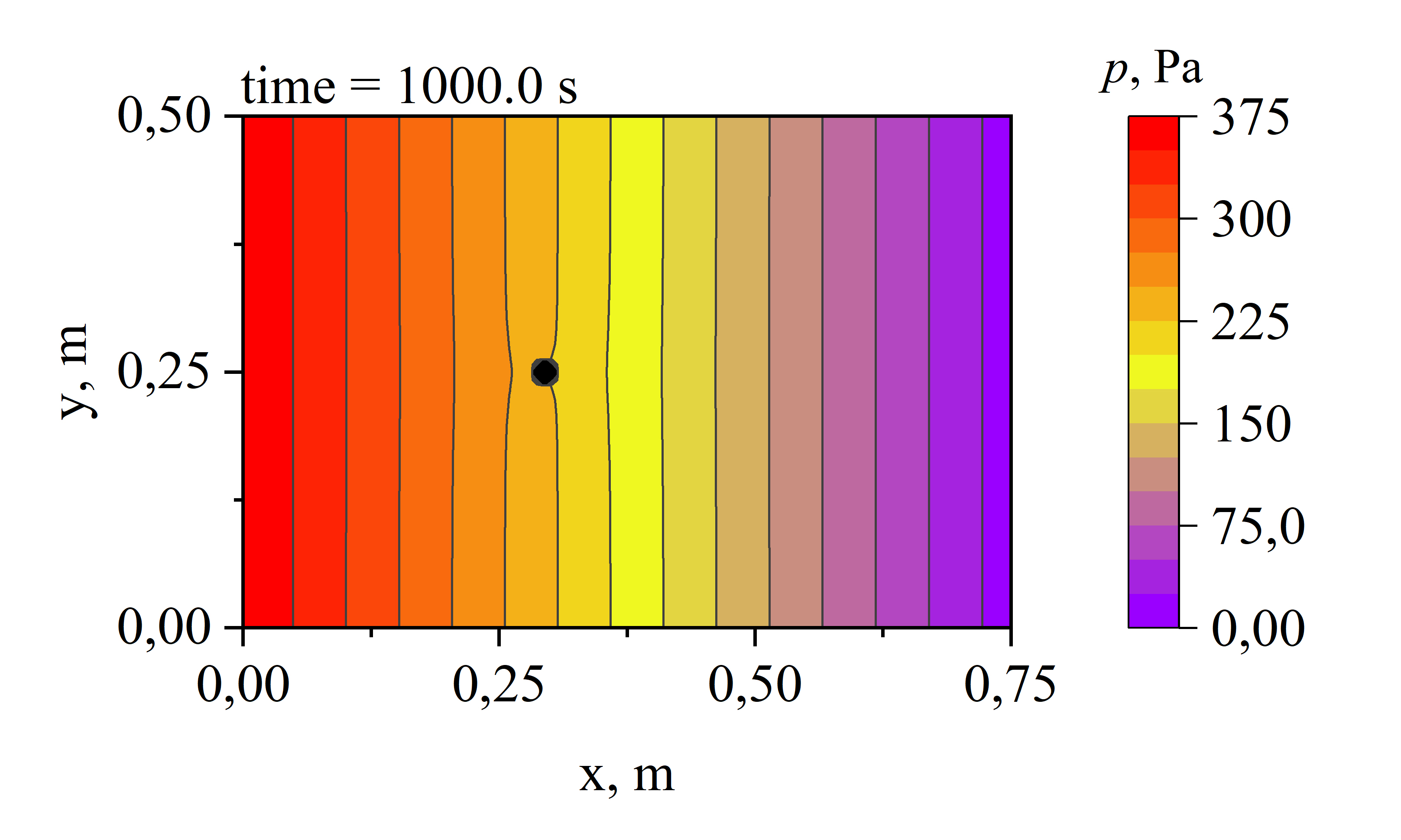}
     \end{subfigure}
          \vfill
    \begin{subfigure}[b]{0.49\textwidth}
         \centering
         \includegraphics[width=0.85\textwidth]{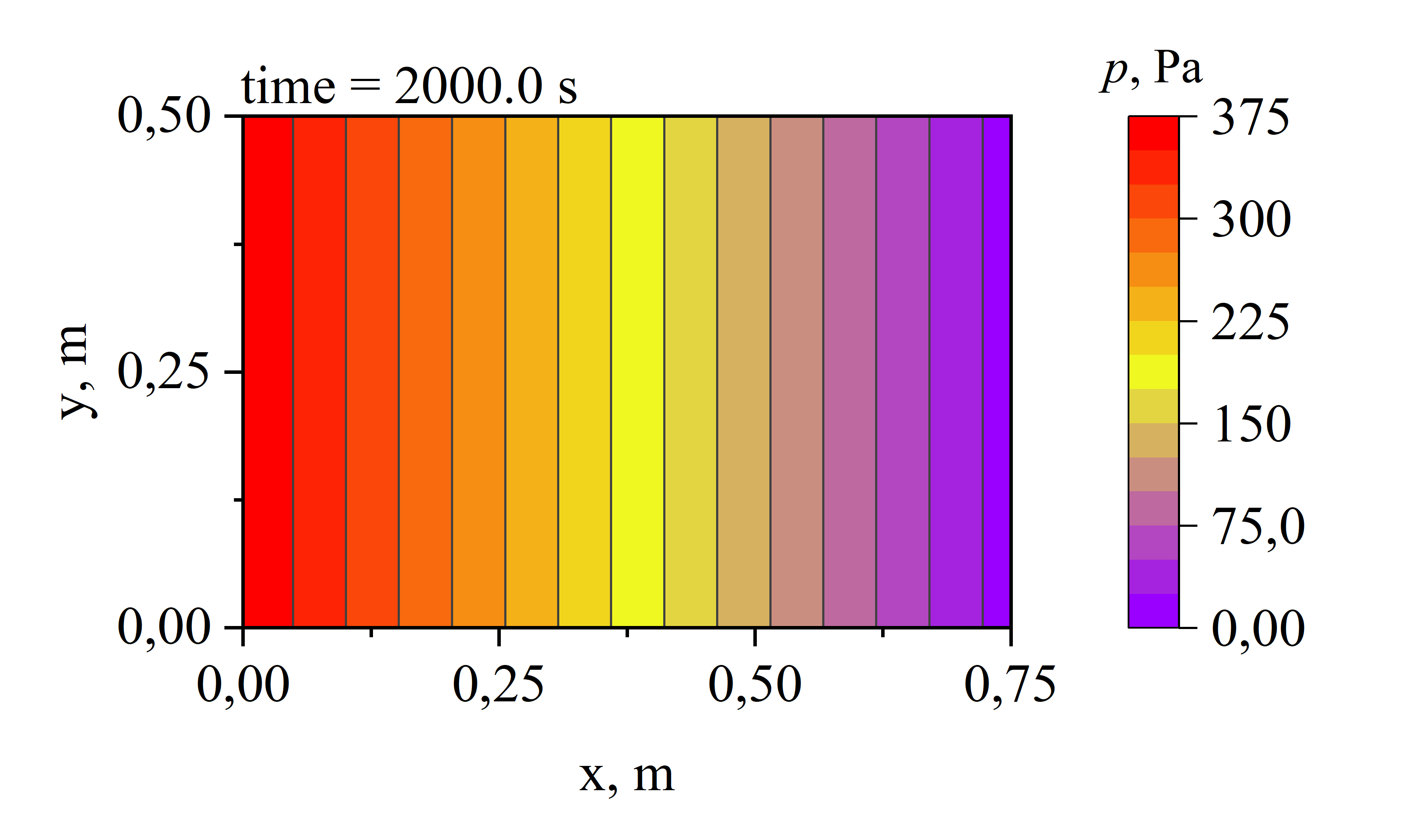}
     \end{subfigure}

    \caption{Pressure distribution, Pa, at different moments of time calculated by the proposed PD model}
    \label{fig:PD_heat_2D_press_distr}
\end{figure}

\begin{figure}[ht]
    \centering
    \includegraphics[width=1\textwidth]{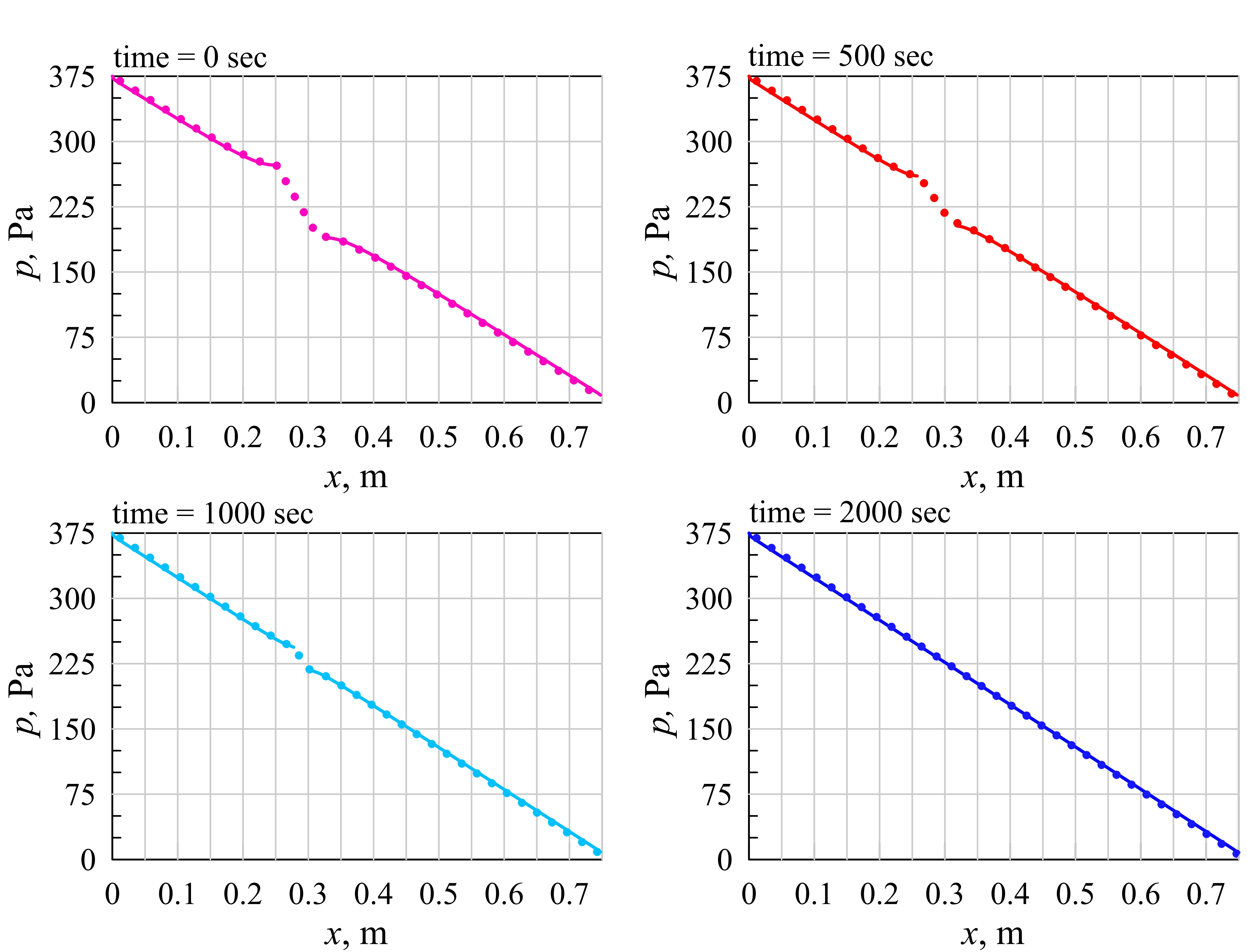}
    \caption{Pressure distribution along the X--X cross-section at different moments of time, calculated by PD and FEM simulations. The solid lines are the developed PD solution; the dots are the analytical solution.}
    \label{fig:PD_heat_2D_press_distr_x}
\end{figure}

The results of PD and FEM simulations are presented in Figs. \ref{fig:PD_heat_2D_temp_distr} -- \ref{fig:PD_heat_2D_press_distr_x}. Firstly, Fig. \ref{fig:PD_heat_2D_temp_distr} shows the temperature distribution in the domain at several time instances. This illustrates how the inclusion is changing size and eventually disappearing, and how the temperature field is deforming due to the water flow.

Comparison between the PD and FEM results is presented in Fig.  \ref{fig:PD_heat_2D_temp_distr_x} (a) for the X--X cross section, (see Fig. \ref{fig:PD_heat_2D}) at time instances 0 s, 500 s, 1000 s, 1500 s, 2000 s, 2500 s and 3000 s. On this figure, we also present the distribution of temperature within the inclusion at the initial time instances 0 s, 50 s, 100 s and 250 s. The results indicate a good agreement between the FEM and PD simulations. Small differences are observed in the temperature distribution within the inclusion as its temperature increases to $T_f$ (results up to time 500 s). This may be explained by the fact that the release of the latent heat of solidification is described differently in the two numerical methods. The temperature distribution for positive temperatures are in very good agreement. Notably, the agreement can be improved by decreasing the particle size and the time step.
 
Our model allows us to determine the pressure distribution within the domain at any moment of time. The results are presented in the plots of Fig. \ref{fig:PD_heat_2D_press_distr} for time instances 0 s, 250 s, 500 s, 1000 s, 2000 s and 3000 s. All ice is melted by 1300 s leading to steady-state pressure field. These figures demonstrate clearly how the pressure field is changing with changing the size of frozen inclusion. The pressure within the inclusion is not calculated, as the water saturation within it provides very low relative permeability, so the t-bonds connected to the inclusion's nodes can be considered as an impermeable solid. 

Comparison between the PD and FEM results for the pressure distribution along the X--X cross-section for the time instances 0 s, 500 s, 1000 s and 2000 s is presented in Fig. \ref{fig:PD_heat_2D_press_distr_x}. These plots provide good evidence of the capability of the PD approach in describing the behaviour of porous material with an evolving discontinuity, as during the simulated period of time, the number of bonds that are participating in water transfer are constantly changing.

The presented agreement between two numerical methods indicates that the non-local PD formulation represents accurately the water flow and heat transfer with phase change in a saturated porous medium. 

It should be noted that the developed PD model can provide a solution with the same high accuracy for the case when all bonds are participating in water transfer. In this case, the model defines the pressure distribution both in the liquid region and within the frozen inclusion for every time instance. As the permeability of the frozen soils is extremely low, the phase change that coincides with density change, occurs under conditions of nearly constant volume that, according to the Clausius–Clapeyron equation, leads to the appearance of extreme negative pressure, see  \cite{Dall_Amico_Endrizzi_Gruber_Rigon_2011, Kurylyk_Watanabe_2013}. This phenomenon is observed in the obtained PD results; however, the FEM model cannot track it at all, see Fig. (\ref{fig:PD_heat_2D_press_distr_x}). The obtained PD pressure values cannot be directly described by the existing forms of the Clausius–Clapeyron equation. However, for the saturated soils with an incompressible solid matrix that are considered in the present study, taking into account this negative pressure does not have a significant physical meaning and does not affect the accuracy of the solution. This effect is out the scope of the present paper. However, in planned future work considering partially saturated soils with deformable soil matrices, the appearance of extreme pressure values and gradients within the frozen soils have to be considered.

Such coupled problems are challenging for the FEM. It is widely accepted that the mesh size and the time step of the model strongly affect the computations and the results. Some geometric model parameters lead to instability of computations and divergent solutions. The FEM solution is particularly sensitive to the flow velocity. This is the reason for our selection of a relatively low pressure gradient in this example. Higher values may lead to computational instability and incorrect results. For example, we could not obtain a converging solution for the present problem for pressures larger than $\approx 3000$ Pa by FEM. In contrast, the PD implementation can effectively deal with such conditions, as will be shown in the next sub-section.

\subsection{Heat transfer with high velocity water flow and phase change}\label{sec:verify_fast}
To illustrate the effectiveness of the developed PD model for the case of convection-dominated heat transfer with phase change, we consider the following example. We suppose that a frozen body is divided into two parts by a narrow unfrozen 'crack'. The initial thickness of this crack is 0.09 m and its shape is given by a $sin$-function with a period 0.5 m and an amplitude 0.15 m. The domain and its boundary conditions are illustrated in Fig. \ref{fig:PD_heat_2D}. The values of the initial and boundary conditions are included in Table \ref{table:inclusion}.

\begin{figure}[ht]
    \centering
    \includegraphics[width=0.7\textwidth]{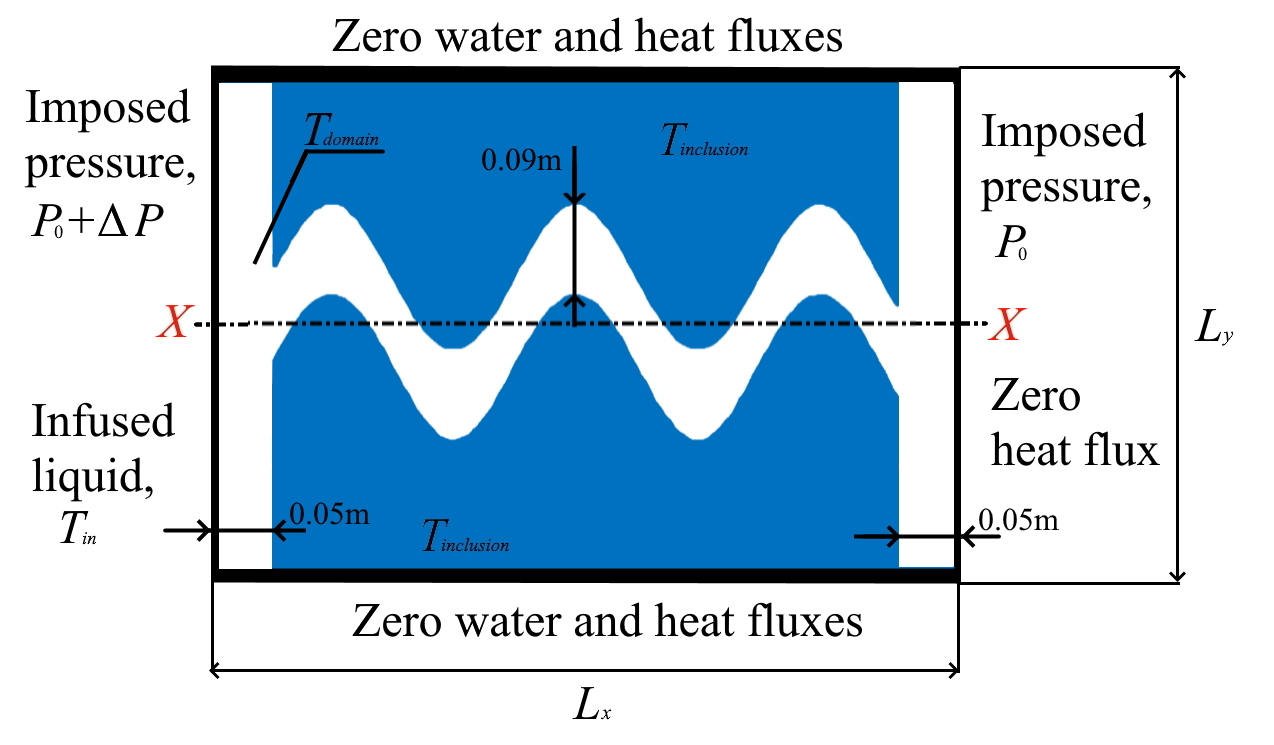}
    \caption{Schematic of an unfrozen 'crack' in a frozen body: geometry, boundary and initial conditions}
    \label{fig:PD_heat_2D_2}
\end{figure}

\begin{figure}[ht]
    \centering
     \begin{subfigure}[b]{0.49\textwidth}
         \centering
         \includegraphics[width=0.85\textwidth]{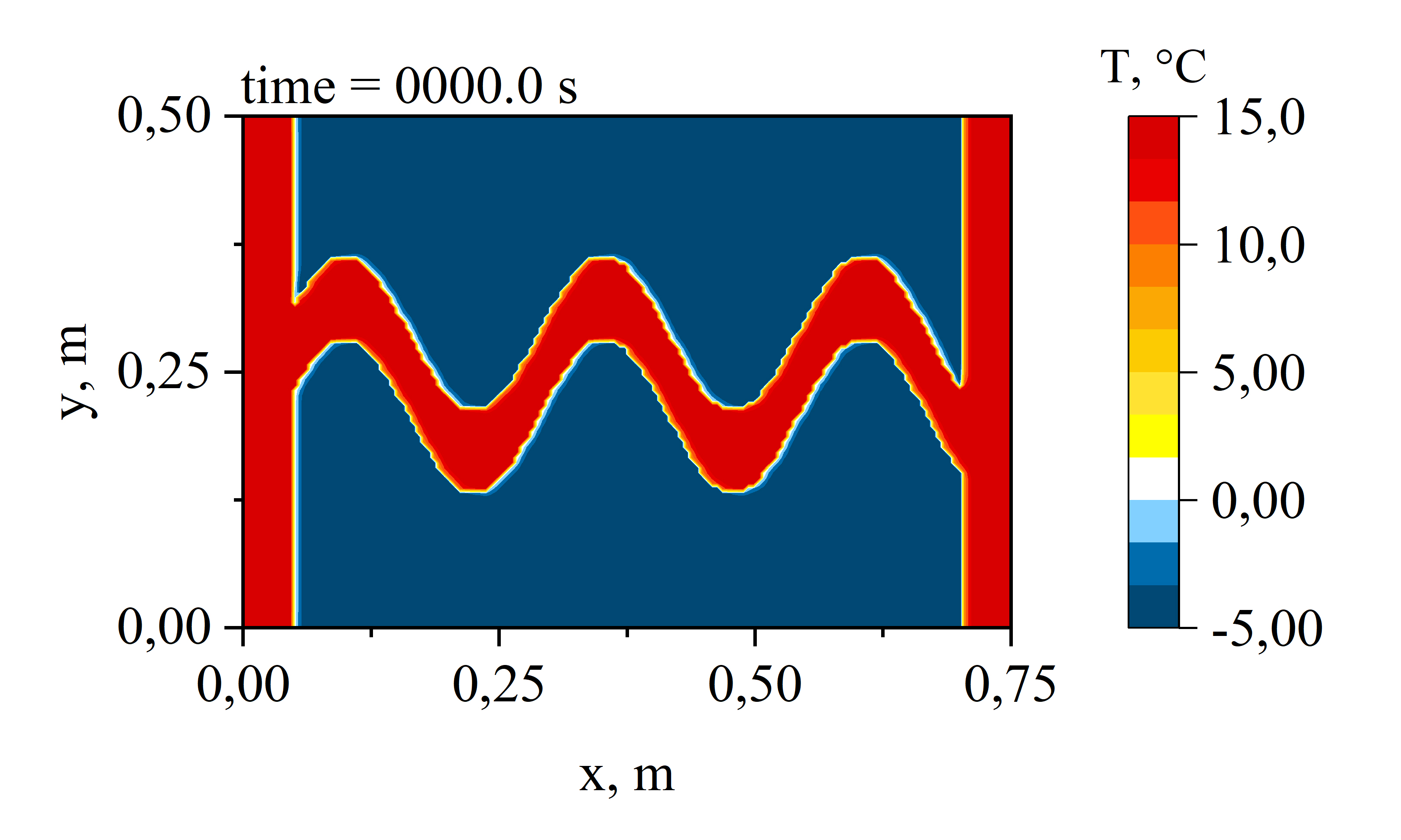}
     \end{subfigure}
     \hfill
     \begin{subfigure}[b]{0.49\textwidth}
         \centering
         \includegraphics[width=0.85\textwidth]{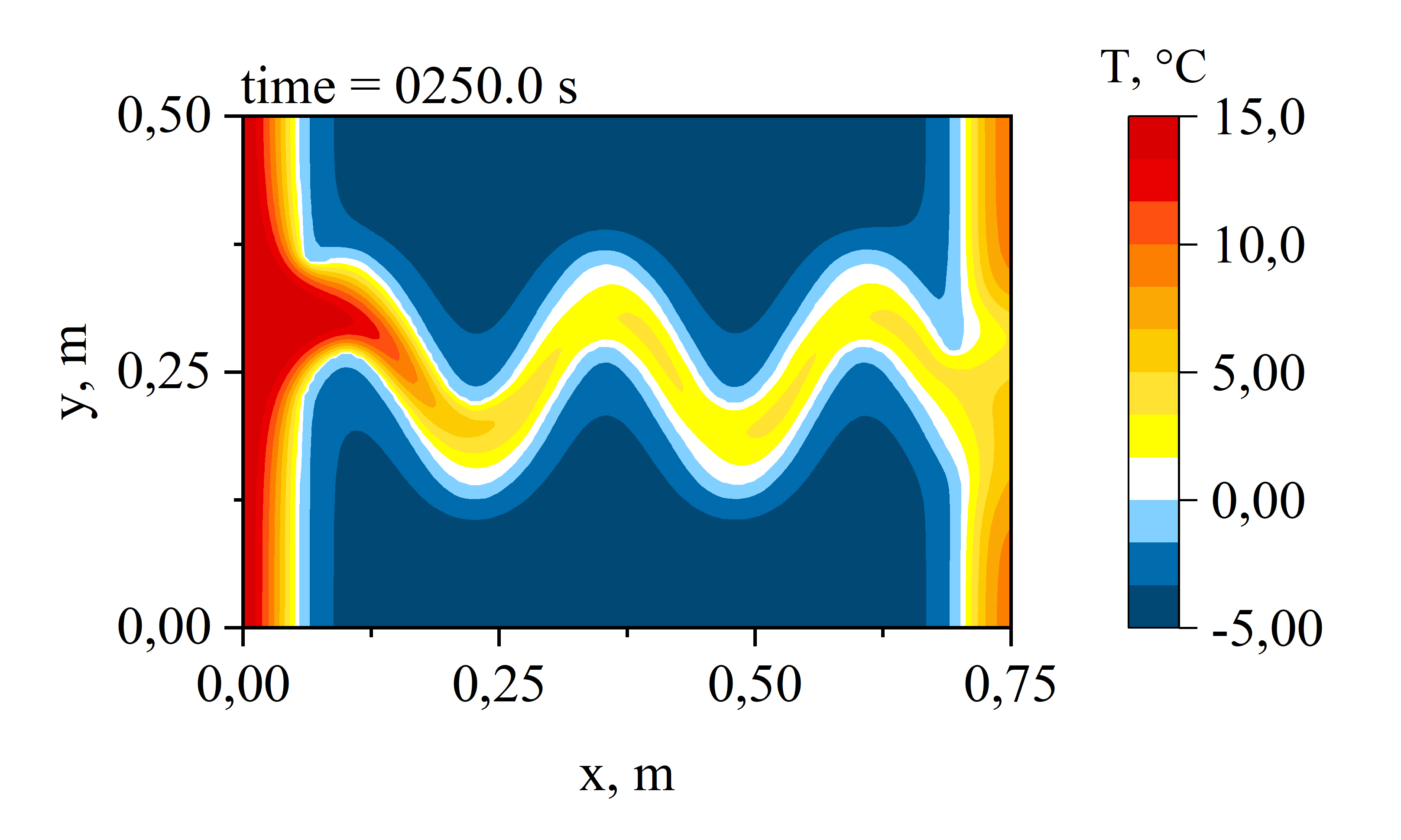}
     \end{subfigure}
     \vfill
    \begin{subfigure}[b]{0.49\textwidth}
         \centering
         \includegraphics[width=0.85\textwidth]{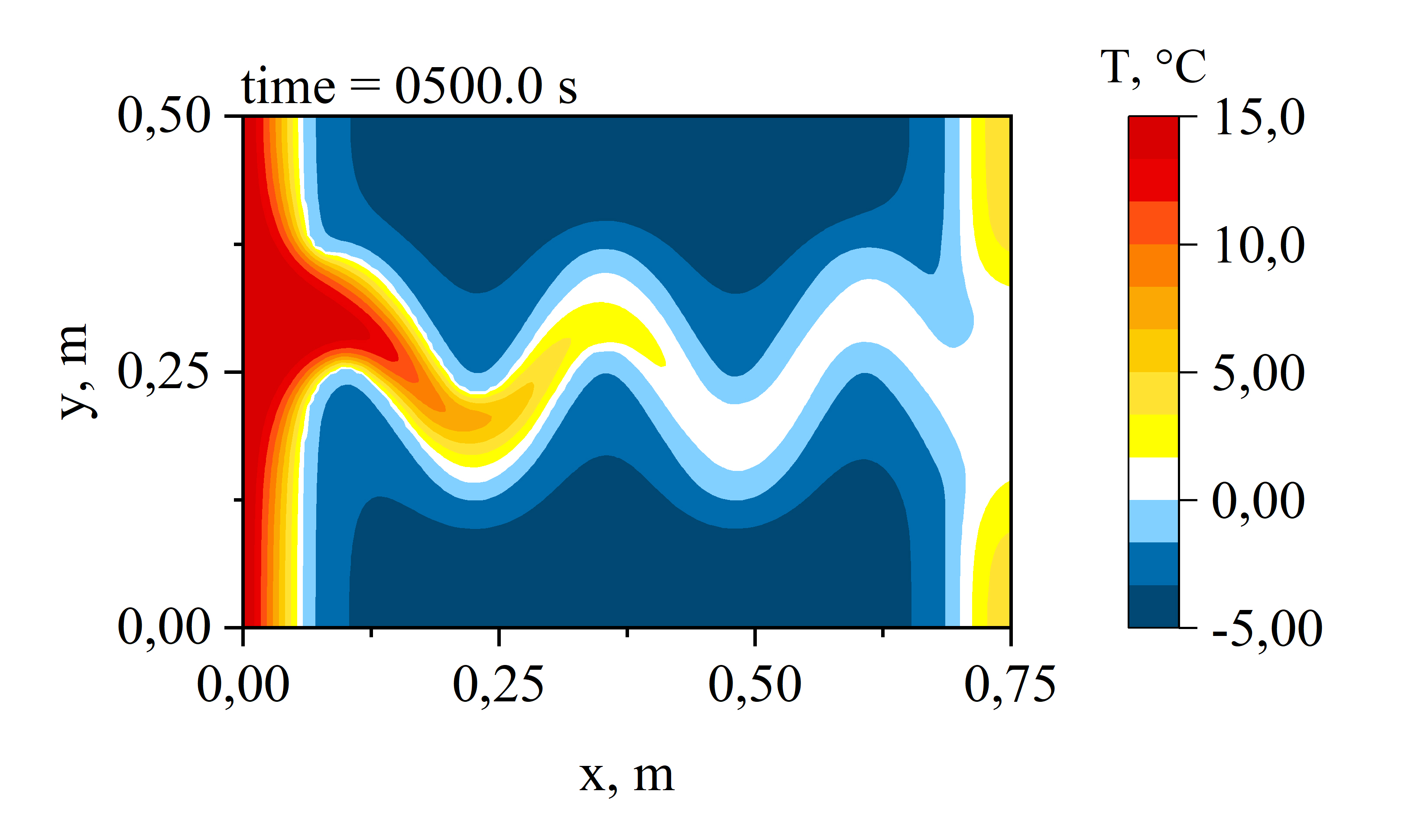}
     \end{subfigure}
     \hfill
     \begin{subfigure}[b]{0.49\textwidth}
         \centering
         \includegraphics[width=0.85\textwidth]{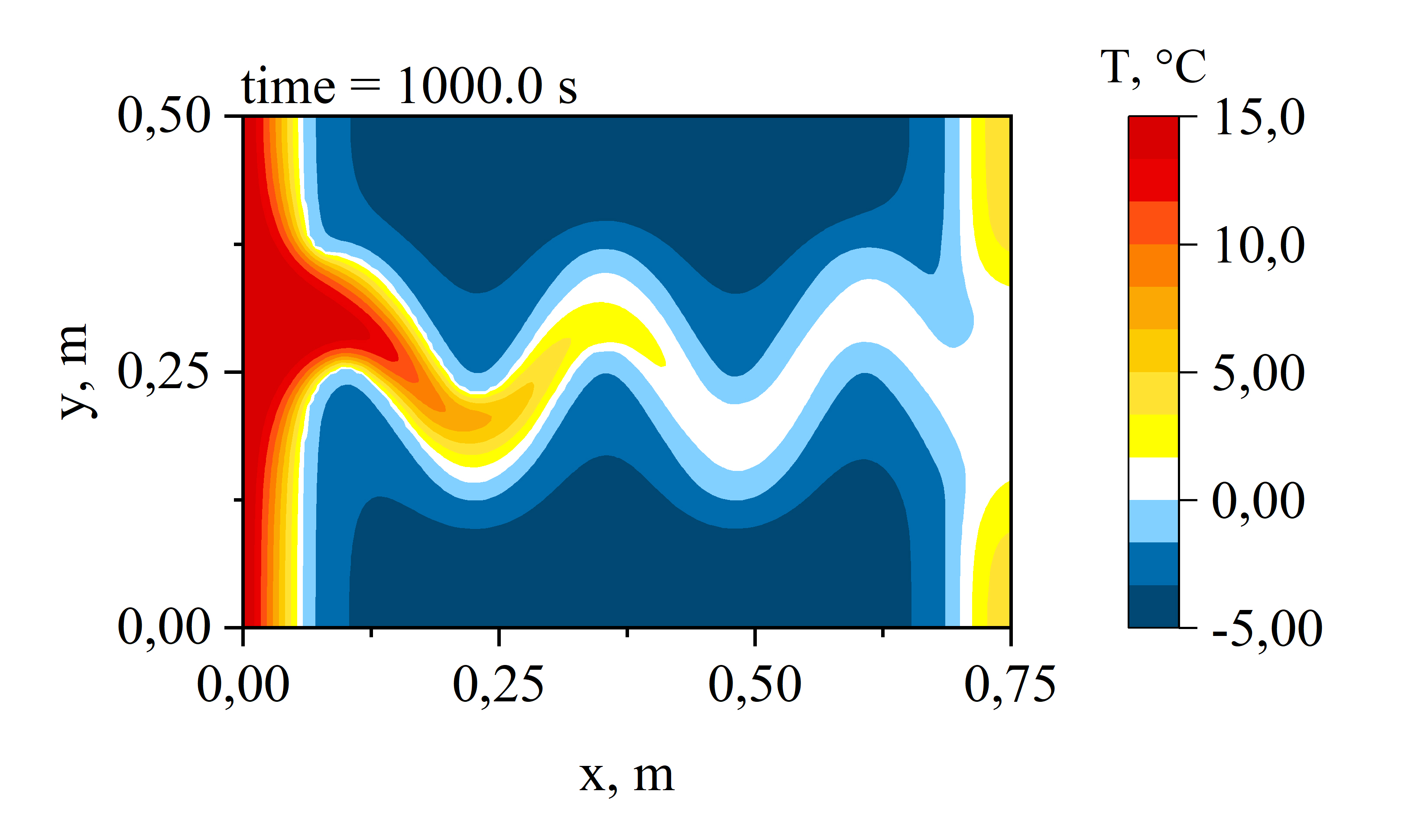}
     \end{subfigure}
     \vfill
    \begin{subfigure}[b]{0.49\textwidth}
         \centering
         \includegraphics[width=0.85\textwidth]{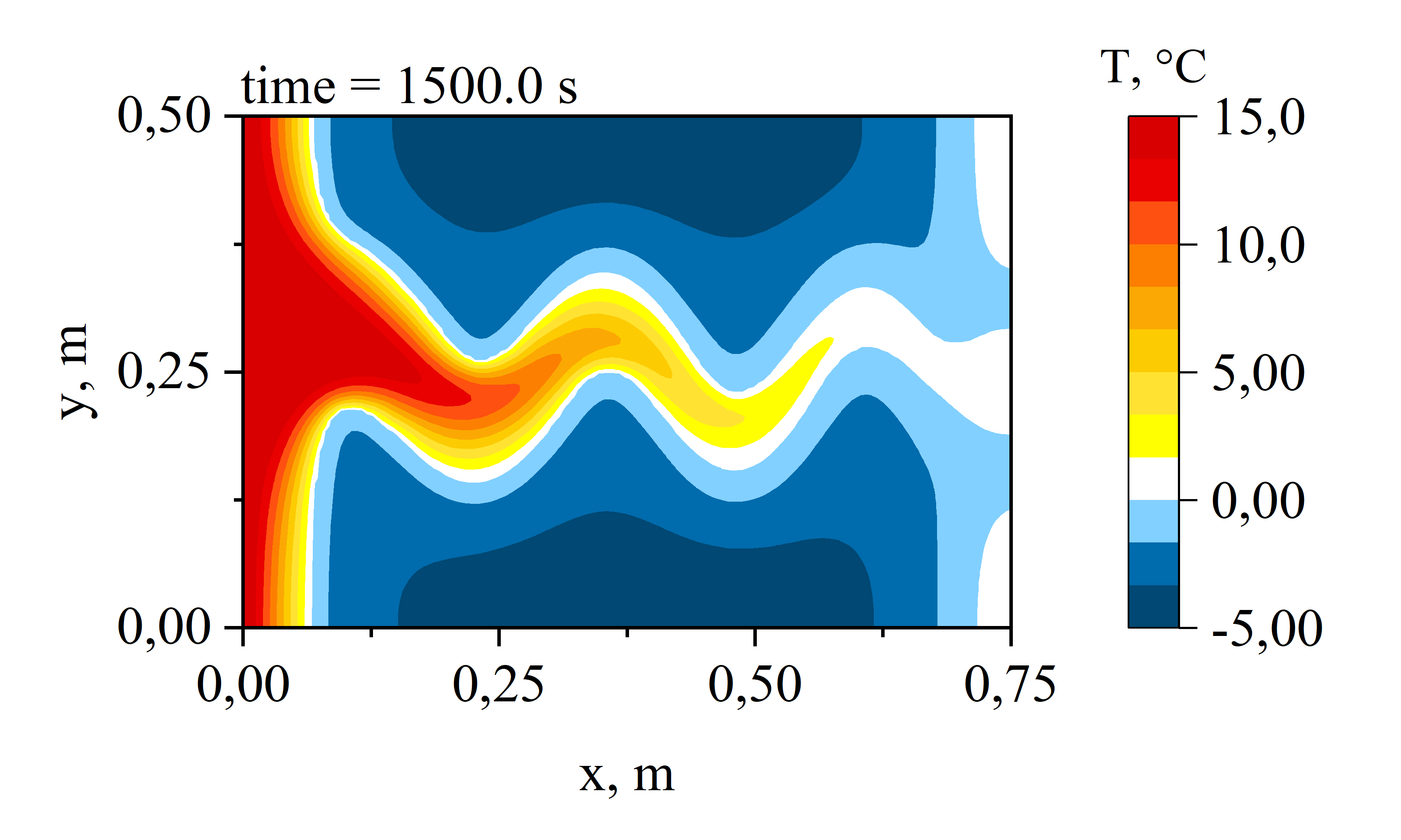}
     \end{subfigure}
          \hfill
     \begin{subfigure}[b]{0.49\textwidth}
         \centering
         \includegraphics[width=0.85\textwidth]{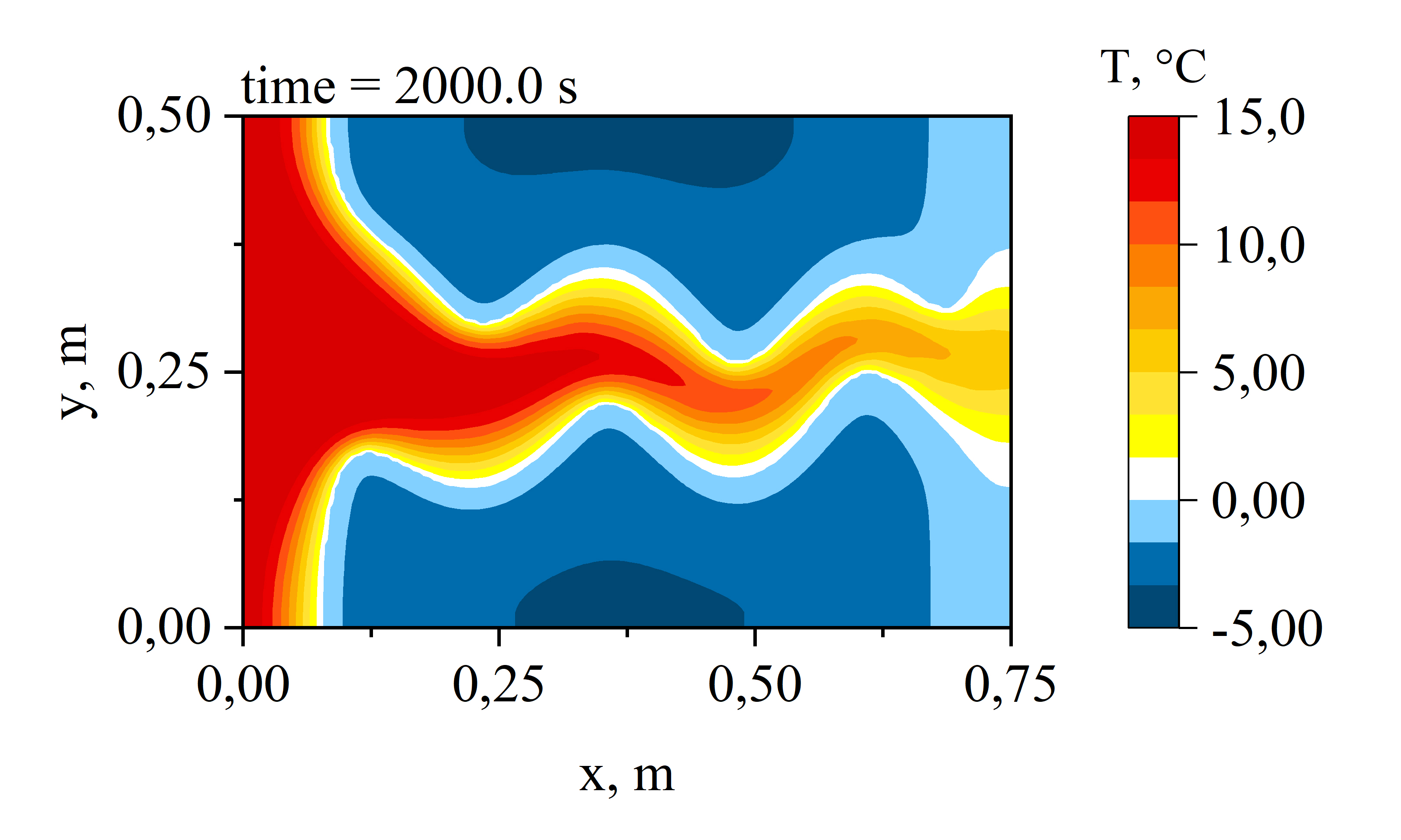}
     \end{subfigure}
    \vfill
    \begin{subfigure}[b]{0.49\textwidth}
         \centering
         \includegraphics[width=0.85\textwidth]{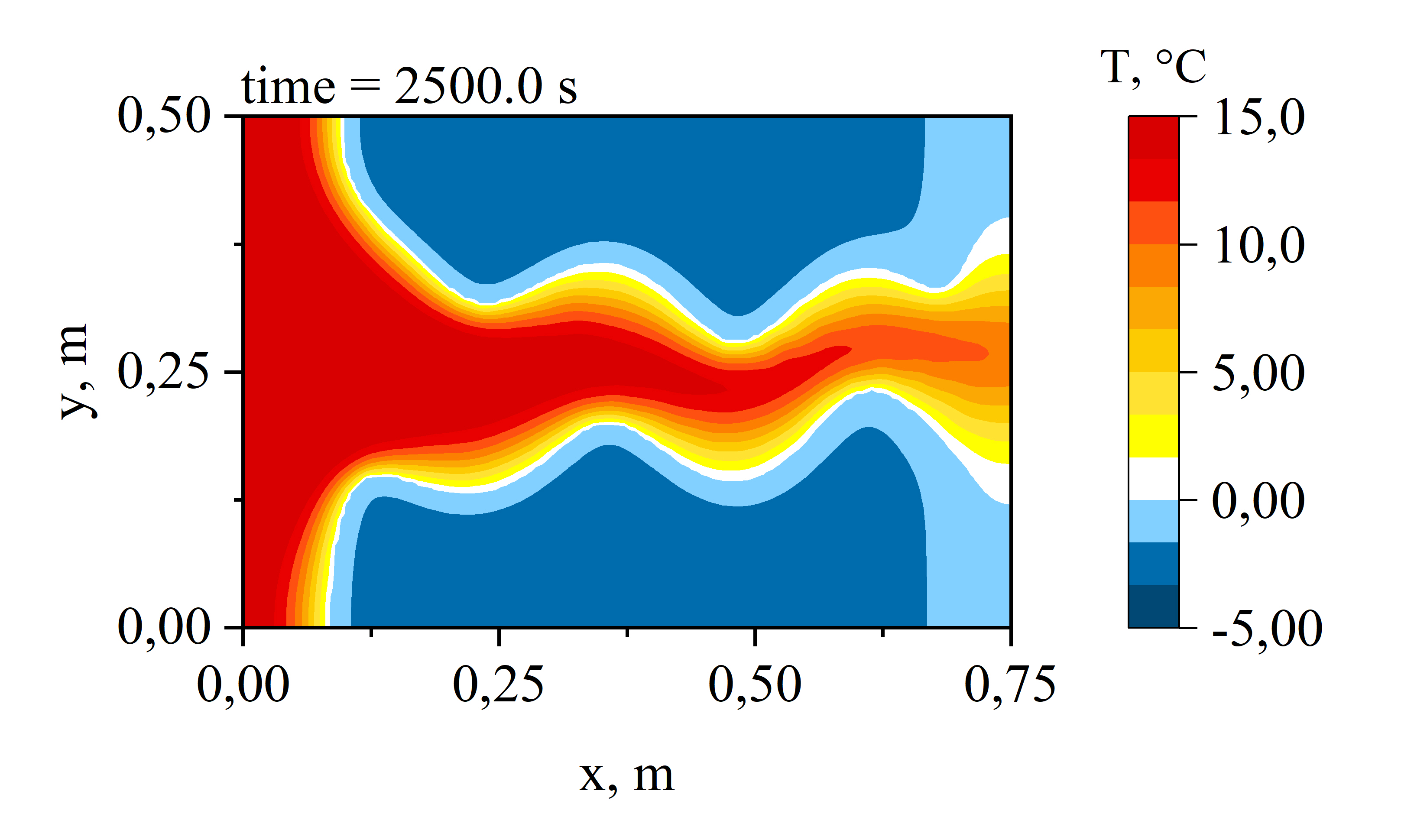}
     \end{subfigure}
               \hfill
     \begin{subfigure}[b]{0.49\textwidth}
         \centering
         \includegraphics[width=0.85\textwidth]{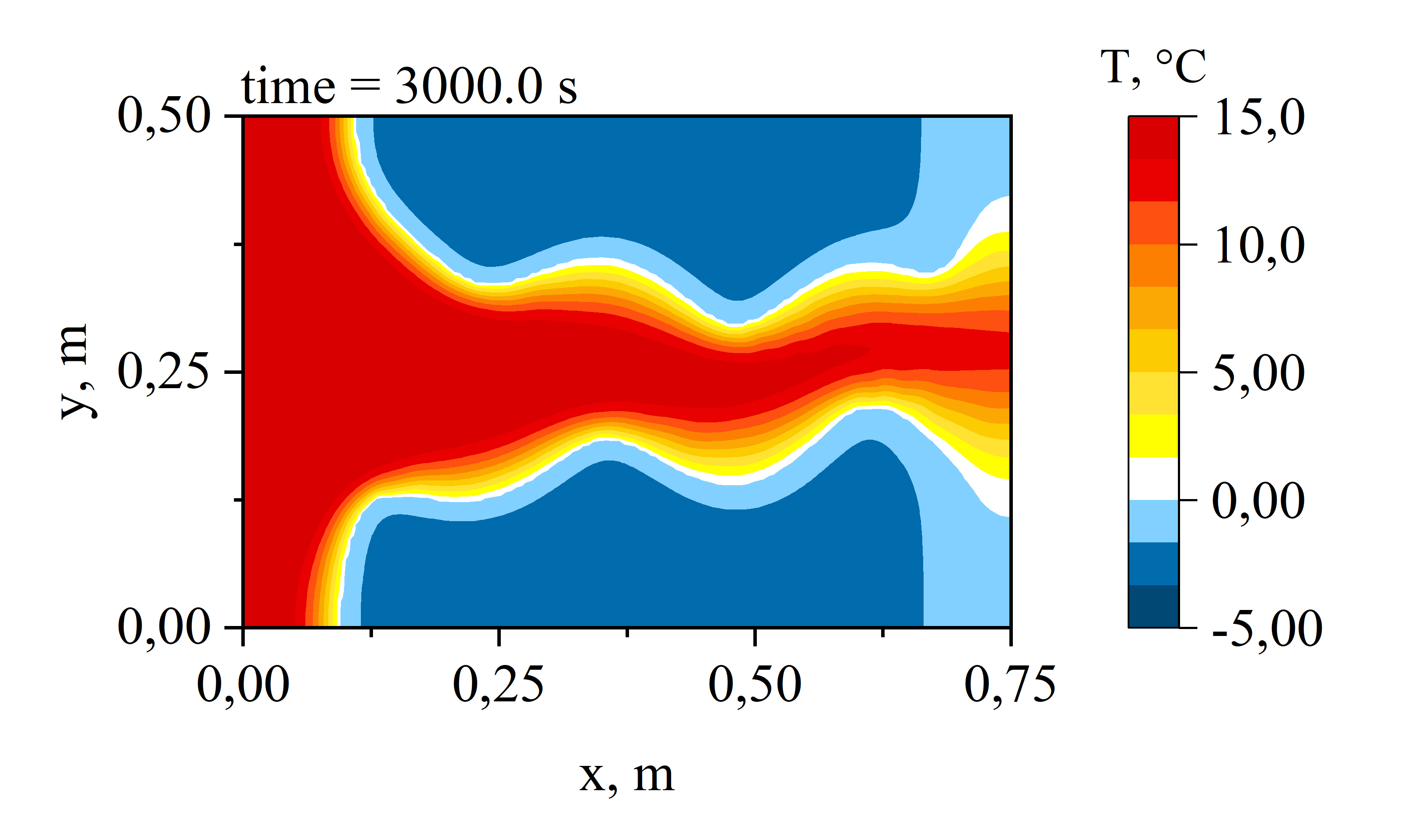}
     \end{subfigure}
    \caption{Temperature distribution, $^\circ {\rm C}$, at different moments of time, according to the proposed PD model}
    \label{fig:PD_heat_2D_temp_distr_4th}
\end{figure}

The PD simulations were performed with particle size $\Delta x=0.005$ m, horizon radius $\delta = 0.015$, and time step $\Delta t=2$ s. The temperature distribution in the domain at several time instances is presented in Fig. \ref{fig:PD_heat_2D_temp_distr_4th}. The figure illustrates how the crack changes its shape and expands due to the heat flux from the high velocity water flow. Initially, the temperature within the 'crack' decreases from its initial value due to heat exchange with the frozen part. The constant inflow of warm liquid increases the temperature within the 'crack' over time and after 2000 s the crack has changed fully. 

This example illustrates the ability of our bond-based PD implementation to describe convection-dominated heat transfer in a porous medium with phase change. It shows clearly how fast inflow of warm water can melt a significant amount of frozen water in a very short period of time - a situation that we found to be very challenging using computational methods based on a local formulation.

\section{Conclusions} \label{sec:conclusions}
This paper presented a non-local formulation, based on bond-based peridynamics, for heat transfer with phase change in saturated porous media under conditions of pressure driven water flow. It allows for the tracking of the interface between phases and for calculating the temperature and pressure distributions within the medium. The model and its numerical implementation were verified by comparison with results obtained using existing analytical solutions for 1D problems, and with transient finite element solutions for a 2D problem. The agreement found by the verification exercises demonstrates the correct implementation of the proposed methodology. 

The detailed description of coupled heat and hydraulic processes is a critical step towards a thermo-hydro-mechanical model, which will allow for example, to describe the hydrological behaviour of permafrost soils and the frost heave phenomenon that greatly affects civil and mining construction works.

\section*{Acknowledgements}\label{Acknowledgements}
The financial support received by Petr Nikolaev in the form of a President doctoral scholarship award (PDS award) by the University of Manchester is gratefully acknowledged.
Jivkov and Margetts are grateful for the financial support of the Engineering and Physical Sciences Research Council, UK (EPSRC) via grant EP/N026136/1.



\bibliographystyle{unsrt}
\bibliography{Main}

@book{Hahn2012HeatConduction,
    title = {{Heat conduction}},
    year = {2012},
    author = {Hahn, David W and {\"{O}}zisik, M Necati},
    edition = {3rd},
    pages = {718},
    publisher = {John Wiley {\&} Sons, Inc.},
    isbn = {1118330110}
}

@article{Silling2000ReformulationForces,
    title = {{Reformulation of elasticity theory for discontinuities and long-range forces}},
    year = {2000},
    journal = {Journal of the Mechanics and Physics of Solids},
    author = {Silling, S.A.},
    number = {1},
    month = {1},
    pages = {175--209},
    volume = {48},
    doi = {10.1016/S0022-5096(99)00029-0},
    issn = {00225096}
}

@article{Bobaru2012ADiscontinuities,
    title = {{A peridynamic formulation for transient heat conduction in bodies with evolving discontinuities}},
    year = {2012},
    journal = {Journal of Computational Physics},
    author = {Bobaru, Florin and Duangpanya, Monchai},
    number = {7},
    pages = {2764--2785},
    volume = {231},
    doi = {10.1016/j.jcp.2011.12.017},
    issn = {10902716}
}

@book{Madenci2014PeridynamicApplications,
    title = {{Peridynamic Theory and Its Applications}},
    year = {2014},
    booktitle = {Peridynamic Theory and Its Applications},
    author = {Madenci, Erdogan and Oterkus, Erkan},
    pages = {1--289},
    volume = {9781461484},
    publisher = {Springer New York},
    address = {New York, NY},
    isbn = {978-1-4614-8464-6},
    doi = {10.1007/978-1-4614-8465-3}
}

@article{Javili2019PeridynamicsReview,
    title = {{Peridynamics review}},
    year = {2019},
    journal = {Mathematics and Mechanics of Solids},
    author = {Javili, Ali and Morasata, Rico and Oterkus, Erkan and Oterkus, Selda},
    number = {11},
    pages = {3714--3739},
    volume = {24},
    doi = {10.1177/1081286518803411},
    issn = {17413028}
}

@article{Liu2020APeridynamics,
    title = {{A new type of peridynamics: Element-based peridynamics}},
    year = {2020},
    journal = {Computer Methods in Applied Mechanics and Engineering},
    author = {Liu, Shuo and Fang, Guodong and Liang, Jun and Fu, Maoqing and Wang, Bing},
    month = {7},
    pages = {113098},
    volume = {366},
    publisher = {Elsevier B.V.},
    doi = {10.1016/j.cma.2020.113098},
    issn = {00457825}
}

@article{Bobaru2010TheConduction,
    title = {{The peridynamic formulation for transient heat conduction}},
    year = {2010},
    journal = {International Journal of Heat and Mass Transfer},
    author = {Bobaru, Florin and Duangpanya, Monchai},
    number = {19-20},
    pages = {4047--4059},
    volume = {53},
    publisher = {Elsevier Ltd},
    doi = {10.1016/j.ijheatmasstransfer.2010.05.024},
    issn = {00179310}
}

@article{Katiyar2014AMedia,
    title = {{A peridynamic formulation of pressure driven convective fluid transport in porous media}},
    year = {2014},
    journal = {Journal of Computational Physics},
    author = {Katiyar, Amit and Foster, John T. and Ouchi, Hisanao and Sharma, Mukul M.},
    pages = {209--229},
    volume = {261},
    publisher = {Elsevier Inc.},
    doi = {10.1016/j.jcp.2013.12.039},
    issn = {10902716}
}

@article{Jabakhanji2015AMedia,
    title = {{A peridynamic model of flow in porous media}},
    year = {2015},
    journal = {Advances in Water Resources},
    author = {Jabakhanji, Rami and Mohtar, Rabi H.},
    number = {January},
    month = {4},
    pages = {22--35},
    volume = {78},
    publisher = {Elsevier Ltd},
    doi = {10.1016/j.advwatres.2015.01.014},
    issn = {03091708}
}

@article{Sedighi2020PeridynamicsErosion,
    title = {{Peridynamics modelling of clay erosion}},
    year = {2020},
    journal = {G{\'{e}}otechnique},
    author = {Sedighi, Majid and Yan, Huaxiang and Jivkov, Andrey P},
    number = {0},
    volume = {0},
    month = {12},
    pages = {1--12},
    doi = {10.1680/jgeot.20.P.149},
    issn = {0016-8505}
}

@article{Zhao2018ConstructionProblems,
    title = {{Construction of a peridynamic model for transient advection-diffusion problems}},
    year = {2018},
    journal = {International Journal of Heat and Mass Transfer},
    author = {Zhao, Jiangming and Chen, Ziguang and Mehrmashhadi, Javad and Bobaru, Florin},
    month = {11},
    pages = {1253--1266},
    volume = {126},
    doi = {10.1016/j.ijheatmasstransfer.2018.06.075},
    issn = {00179310}
}

@article{Yan2020PeridynamicsMedia,
    title = {{Peridynamics modelling of coupled water flow and chemical transport in unsaturated porous media}},
    year = {2020},
    journal = {Journal of Hydrology},
    author = {Yan, Huaxiang and Sedighi, Majid and Jivkov, Andrey P.},
    number = {October},
    volume = {591},
    pages = {125648},
    publisher = {Elsevier B.V.},
    doi = {10.1016/j.jhydrol.2020.125648},
    issn = {00221694}
}

@article{Rowland2011ThePermafrost,
    title = {{The role of advective heat transport in talik development beneath lakes and ponds in discontinuous permafrost}},
    year = {2011},
    journal = {Geophysical Research Letters},
    author = {Rowland, J.C. C. and Travis, B.J. J. and Wilson, C.J. J.},
    number = {17},
    month = {9},
    pages = {n/a-n/a},
    volume = {38},
    doi = {10.1029/2011GL048497},
    issn = {00948276}
}

@article{McKenzie2007GroundwaterBogs,
    title = {{Groundwater flow with energy transport and water–ice phase change: Numerical simulations, benchmarks, and application to freezing in peat bogs}},
    year = {2007},
    journal = {Advances in Water Resources},
    author = {McKenzie, Jeffrey M. and Voss, Clifford I. and Siegel, Donald I.},
    number = {4},
    month = {4},
    pages = {966--983},
    volume = {30},
    doi = {10.1016/j.advwatres.2006.08.008},
    issn = {03091708}
}

@article{Frederick2014TaliksConditions,
    title = {{Taliks in relict submarine permafrost and methane hydrate deposits: Pathways for gas escape under present and future conditions}},
    year = {2014},
    journal = {Journal of Geophysical Research: Earth Surface},
    author = {Frederick, J M and Buffett, B A},
    number = {2},
    month = {2},
    pages = {106--122},
    volume = {119},
    doi = {10.1002/2013JF002987},
    issn = {21699003}
}

@article{Vidstrand2013ModelingSweden,
    title = {{Modeling of groundwater flow at depth in crystalline rock beneath a moving ice-sheet margin, exemplified by the Fennoscandian Shield, Sweden}},
    year = {2013},
    journal = {Hydrogeology Journal},
    author = {Vidstrand, Patrik and Follin, Sven and Selroos, Jan-Olof and N{\"{a}}slund, Jens-Ove and Rh{\'{e}}n, Ingvar},
    number = {1},
    month = {2},
    pages = {239--255},
    volume = {21},
    isbn = {1004001209218},
    doi = {10.1007/s10040-012-0921-8},
    issn = {1431-2174}
}

@article{Grenier2013ImpactSystem,
    title = {{Impact of permafrost development on groundwater flow patterns: a numerical study considering freezing cycles on a two-dimensional vertical cut through a generic river-plain system}},
    year = {2013},
    journal = {Hydrogeology Journal},
    author = {Grenier, Christophe and R{\'{e}}gnier, Damien and Mouche, Emmanuel and Benabderrahmane, Hakim and Costard, François and Davy, Philippe},
    number = {1},
    month = {2},
    pages = {257--270},
    volume = {21},
    doi = {10.1007/s10040-012-0909-4},
    issn = {1431-2174}
}

@article{Vitel2016,
    title = {{Modeling heat and mass transfer during ground freezing subjected to high seepage velocities}},
    year = {2016},
    journal = {Computers and Geotechnics},
    author = {Vitel, M. and Rouabhi, A. and Tijani, M. and Gu{\'{e}}rin, F.},
    pages = {1--15},
    volume = {73},
    publisher = {Elsevier Ltd},
    doi = {10.1016/j.compgeo.2015.11.014},
    issn = {18737633}
}

@article{Zhou2013,
    title = {{A three-phase thermo-hydro-mechanical finite element model for freezing soils}},
    year = {2013},
    journal = {International Journal for Numerical and Analytical Methods in Geomechanics},
    author = {Zhou, M. M. and Meschke, G.},
    number = {18},
    pages = {3173--3193},
    volume = {37},
    publisher = {Wiley Online Library},
    doi = {10.1002/nag.2184},
    issn = {03639061}
}

@article{Tounsi2020Thermo-hydro-mechanicalFluid,
    title = {{Thermo-hydro-mechanical modeling of artificial ground freezing taking into account the salinity of the saturating fluid}},
    year = {2020},
    journal = {Computers and Geotechnics},
    author = {Tounsi, H. and Rouabhi, A. and Jahangir, E.},
    number = {September 2019},
    month = {3},
    pages = {103382},
    volume = {119},
    publisher = {Elsevier},
    doi = {10.1016/j.compgeo.2019.103382},
    issn = {0266352X}
}

@article{Zueter2020ThermalModeling,
    title = {{Thermal and hydraulic analysis of selective artificial ground freezing using air insulation: Experiment and modeling}},
    year = {2020},
    journal = {Computers and Geotechnics},
    author = {Zueter, Ahmad and Nie-Rouquette, Aurelien and Alzoubi, Mahmoud A. and Sasmito, Agus P.},
    number = {July 2019},
    month = {4},
    pages = {103416},
    volume = {120},
    publisher = {Elsevier},
    doi = {10.1016/j.compgeo.2019.103416},
    issn = {0266352X}
}

@article{Oterkus2017FullyFractures,
    title = {{Fully coupled poroelastic peridynamic formulation for fluid-filled fractures}},
    year = {2017},
    journal = {Engineering Geology},
    author = {Oterkus, Selda and Madenci, Erdogan and Oterkus, Erkan},
    month = {7},
    pages = {19--28},
    volume = {225},
    doi = {10.1016/j.enggeo.2017.02.001},
    issn = {00137952}
}

@article{Chen2015SelectingDiffusion,
    title = {{Selecting the kernel in a peridynamic formulation: A study for transient heat diffusion}},
    year = {2015},
    journal = {Computer Physics Communications},
    author = {Chen, Ziguang and Bobaru, Florin},
    pages = {51--60},
    volume = {197},
    publisher = {Elsevier B.V.},
    doi = {10.1016/j.cpc.2015.08.006},
    issn = {00104655}
}

@article{Grenier2018GroundwaterCases,
    title = {{Groundwater flow and heat transport for systems undergoing freeze-thaw: Intercomparison of numerical simulators for 2D test cases}},
    year = {2018},
    journal = {Advances in Water Resources},
    author = {Grenier, Christophe and Anbergen, Hauke and Bense, Victor and Chanzy, Quentin and Coon, Ethan and Collier, Nathaniel and Costard, François and Ferry, Michel and Frampton, Andrew and Frederick, Jennifer and Gon{\c{c}}alv{\`{e}}s, Julio and Holm{\'{e}}n, Johann and Jost, Anne and Kokh, Samuel and Kurylyk, Barret and McKenzie, Jeffrey and Molson, John and Mouche, Emmanuel and Orgogozo, Laurent and Pannetier, Romain and Rivi{\`{e}}re, Agnès and Roux, Nicolas and R{\"{u}}haak, Wolfram and Scheidegger, Johanna and Selroos, Jan Olof and Therrien, René and Vidstrand, Patrik and Voss, Clifford},
    number = {January},
    pages = {196--218},
    volume = {114},
    publisher = {Elsevier Ltd},
    doi = {10.1016/j.advwatres.2018.02.001},
    issn = {03091708}
}

@article{Madenci2017,
    author = {Madenci, Erdogan and Dorduncu, Mehmet and Barut, Atila and Futch, Michael},
    doi = {10.1002/num.22167},
    issn = {0749159X},
    journal = {Numerical Methods for Partial Differential Equations},
    month = {sep},
    number = {5},
    pages = {1726--1753},
    title = {{Numerical solution of linear and nonlinear partial differential equations using the peridynamic differential operator}},
    volume = {33},
    year = {2017}
}

@book{Madenci2019,
    address = {Cham},
    author = {Madenci, Erdogan and Barut, Atila and Dorduncu, Mehmet},
    doi = {10.1007/978-3-030-02647-9},
    isbn = {978-3-030-02646-2},
    publisher = {Springer International Publishing},
    title = {{Peridynamic Differential Operator for Numerical Analysis}},
    year = {2019}
}

@article{Wang2018,
    author = {Wang, Linjuan and Xu, Jifeng and Wang, Jianxiang},
    doi = {10.1016/j.ijheatmasstransfer.2017.11.074},
    issn = {00179310},
    journal = {International Journal of Heat and Mass Transfer},
    month = {mar},
    pages = {1284--1292},
    publisher = {Elsevier Ltd},
    title = {{A peridynamic framework and simulation of non-Fourier and nonlocal heat conduction}},
    volume = {118},
    year = {2018}
}

@article{Madenci2016,
    author = {Madenci, Erdogan and Barut, Atila and Futch, Michael},
    doi = {10.1016/j.cma.2016.02.028},
    issn = {00457825},
    journal = {Computer Methods in Applied Mechanics and Engineering},
    month = {jun},
    pages = {408--451},
    publisher = {Elsevier Ltd},
    title = {{Peridynamic differential operator and its applications}},
    volume = {304},
    year = {2016}
}

@article{LIU2021108504,
title = {Studies on model-scale ice using micro-potential-based peridynamics},
journal = {Ocean Engineering},
volume = {221},
pages = {108504},
year = {2021},
issn = {0029-8018},
doi = {https://doi.org/10.1016/j.oceaneng.2020.108504},
author = {Renwei Liu and Yanzhuo Xue and Duanfeng Han and Baoyu Ni},
}

@article{XUE2020107853,
title = {A review for numerical simulation methods of ship–ice interaction},
journal = {Ocean Engineering},
volume = {215},
pages = {107853},
year = {2020},
issn = {0029-8018},
doi = {https://doi.org/10.1016/j.oceaneng.2020.107853},
author = {Yanzhuo Xue and Renwei Liu and Zheng Li and Duanfeng Han},
}

@article{CHEN2021107463,
title = {A refined thermo-mechanical fully coupled peridynamics with application to concrete cracking},
journal = {Engineering Fracture Mechanics},
volume = {242},
pages = {107463},
year = {2021},
issn = {0013-7944},
doi = {https://doi.org/10.1016/j.engfracmech.2020.107463},
author = {Wanhan Chen and Xin Gu and Qing Zhang and Xiaozhou Xia}
}

@article{CHEN2021104203,
title = {A coupled mechano-chemical peridynamic model for pit-to-crack transition in stress-corrosion cracking},
author={Chen, Ziguang and Jafarzadeh, Siavash and Zhao, Jiangming and Bobaru, Florin},
journal = {Journal of the Mechanics and Physics of Solids},
volume = {146},
pages = {104203},
year = {2021},
issn = {0022-5096},
doi = {https://doi.org/10.1016/j.jmps.2020.104203},
}

@article{ZHAO2020106969,
title = {A stochastic multiscale peridynamic model for corrosion-induced fracture in reinforced concrete},
journal = {Engineering Fracture Mechanics},
volume = {229},
pages = {106969},
year = {2020},
issn = {0013-7944},
doi = {https://doi.org/10.1016/j.engfracmech.2020.106969},
author = {Jiangming Zhao and Ziguang Chen and Javad Mehrmashhadi and Florin Bobaru}
}

@article{CHEN2019104059,
author={Chen, Ziguang and Niazi, Sina and Bobaru, Florin},
title = {A peridynamic model for brittle damage and fracture in porous materials},
journal = {International Journal of Rock Mechanics and Mining Sciences},
volume = {122},
pages = {104059},
year = {2019},
issn = {1365-1609},
doi = {https://doi.org/10.1016/j.ijrmms.2019.104059},
}

@article{Gu2019,
author = {Gu, Xin and Zhang, Qing and Madenci, Erdogan},
doi = {10.1108/EC-09-2018-0433},
issn = {0264-4401},
journal = {Engineering Computations},
month = {oct},
number = {8},
pages = {2557--2587},
title = {{Refined bond-based peridynamics for thermal diffusion}},
volume = {36},
year = {2019}
}

@article{silling2007peridynamic,
  title={Peridynamic states and constitutive modeling},
  author={Silling, Stewart A and Epton, M and Weckner, O and Xu, Ji and Askari, E23481501120},
  journal={Journal of Elasticity},
  volume={88},
  number={2},
  pages={151--184},
  year={2007},
  publisher={Springer}
}

@article{gerstle2008peridynamic,
  title={Peridynamic simulation of electromigration},
  author={Gerstle, Walter and Silling, Stewart and Read, David and Tewary, Vinod and Lehoucq, Richard},
  journal={Comput Mater Continua},
  volume={8},
  number={2},
  pages={75--92},
  year={2008}
}

@article{Agwai2011,
author = {Agwai, Abigail and Guven, Ibrahim and Madenci, Erdogan},
doi = {10.1007/s10704-011-9628-4},
issn = {0376-9429},
journal = {International Journal of Fracture},
month = {sep},
number = {1},
pages = {65--78},
title = {{Predicting crack propagation with peridynamics: a comparative study}},
volume = {171},
year = {2011}
}

@article{Jafari2017,
author = {Jafari, Akbar and Bahaaddini, Reza and Jahanbakhsh, Hasan},
title = {Numerical analysis of peridynamic and classical models in transient heat transfer, employing Galerkin approach},
journal = {Heat Transfer—Asian Research},
volume = {47},
number = {3},
pages = {531-555},
doi = {https://doi.org/10.1002/htj.21317},
year = {2018}
}

@article{Wang2016TheGreen,
author = {Wang, L. J.  and Xu, J. F.  and Wang, J. X. },
title = {The Green's functions for peridynamic non-local diffusion},
journal = {Proceedings of the Royal Society A: Mathematical, Physical and Engineering Sciences},
volume = {472},
number = {2193},
pages = {20160185},
year = {2016},
doi = {10.1098/rspa.2016.0185},
}

@article{WEN2020102927,
title = {Evaluation of the calculated unfrozen water contents determined by different measured subzero temperature ranges},
journal = {Cold Regions Science and Technology},
volume = {170},
pages = {102927},
year = {2020},
issn = {0165-232X},
doi = {https://doi.org/10.1016/j.coldregions.2019.102927},
author = {Haiyan Wen and Jun Bi and Ding Guo},
}

@article{Lu2020Peridynamic,
author = {Lu, Wei and Li, Mingyang and Vazic, Bozo and Oterkus, Selda and Oterkus, Erkan and Wang, Qing},
doi = {10.1017/jmech.2019.61},
issn = {18118216},
journal = {Journal of Mechanics},
number = {2},
pages = {223--234},
title = {{Peridynamic Modelling of Fracture in Polycrystalline Ice}},
volume = {36},
year = {2020}
}

@article{Orgogozo2019,
author = {Orgogozo, Laurent and Prokushkin, Anatoly S. and Pokrovsky, Oleg S. and Grenier, Christophe and Quintard, Michel and Viers, Jérôme and Audry, Stéphane},
title = {Water and energy transfer modeling in a permafrost-dominated, forested catchment of Central Siberia: The key role of rooting depth},
journal = {Permafrost and Periglacial Processes},
volume = {30},
number = {2},
pages = {75-89},
doi = {https://doi.org/10.1002/ppp.1995},
year = {2019}
}

@article{Wan2021,
author = {Wan, Hanli and Bian, Jianmin and Zhang, Han and Li, Yihan},
doi = {10.1007/s11783-020-1302-5},
issn = {2095-2201},
journal = {Frontiers of Environmental Science {\&} Engineering},
month = {feb},
number = {1},
pages = {10},
title = {{Assessment of future climate change impacts on water-heat-salt migration in unsaturated frozen soil using CoupModel}},
volume = {15},
year = {2021}
}

@article{ZHANG2021125603,
title = {Coupling analysis of the heat-water dynamics and frozen depth in a seasonally frozen zone},
journal = {Journal of Hydrology},
volume = {593},
pages = {125603},
year = {2021},
issn = {0022-1694},
doi = {https://doi.org/10.1016/j.jhydrol.2020.125603},
author = {Xudong Zhang and Yajun Wu and Encheng Zhai and Peng Ye},
}

@article{YU2019150,
title = {Comparative study of relating equations in coupled thermal-hydraulic finite element analyses},
journal = {Cold Regions Science and Technology},
volume = {161},
pages = {150-158},
year = {2019},
issn = {0165-232X},
doi = {https://doi.org/10.1016/j.coldregions.2019.03.005},
author = {Zheng Yu and Longcai Yang and Shunhua Zhou},
}

@article{TENG2021125885,
title = {Generalising the Kozeny-Carman equation to frozen soils},
journal = {Journal of Hydrology},
volume = {594},
pages = {125885},
year = {2021},
issn = {0022-1694},
doi = {https://doi.org/10.1016/j.jhydrol.2020.125885},
author = {Jidong Teng and Han Yan and Sihao Liang and Sheng Zhang and Daichao Sheng},

}

@article{Kurylyk_Watanabe_2013, 
 title={The mathematical representation of freezing and thawing processes in variably-saturated, non-deformable soils}, 
 volume={60}, 
 ISSN={03091708},
 DOI={10.1016/j.advwatres.2013.07.016}, 
 journal={Advances in Water Resources}, 
 author={Kurylyk, Barret L. and Watanabe, Kunio}, 
 year={2013}, 
 month={Oct}, 
 pages={160–177} 
 }

@article{Dall_Amico_Endrizzi_Gruber_Rigon_2011, 
title={A robust and energy-conserving model of freezing variably-saturated soil}, 
volume={5}, 
ISSN={1994-0424}, 
DOI={10.5194/tc-5-469-2011}, 
number={2}, 
journal={The Cryosphere}, 
author={Dall’Amico, M. and Endrizzi, S. and Gruber, S. and Rigon, R.}, 
year={2011}, 
month={Jun}, 
pages={469–484}
}

\end{document}